\documentclass[journal]{IEEEtran}  
\usepackage{mathtools}
\usepackage{tikz}

\DeclareMathOperator*{\argmax}{arg\,max}
\DeclareMathOperator*{\argmin}{arg\,min}
\DeclareMathOperator{\diag}{diag}
\DeclareMathOperator*{\minimize}{minimize}

\renewcommand{\Pr}{\textnormal{Pr}}
\newcommand\E{\mathbb{E}}
\newcommand{\trans}{\ensuremath{^{\mathsf{T}}}}

\definecolor{myblue}{RGB}{0,100,200}

\usepackage{moreverb}
\usepackage{epsfig}
\usepackage{amsmath,amssymb,amsthm,mathrsfs,amsfonts,dsfont}
\usepackage{tikz}
\usepackage{fancyhdr}
\usepackage{adjustbox,lipsum}
\usepackage[font={small}]{caption}
\usepackage{color,soul}
\usepackage{amsthm,amssymb,amsmath,bm}
\usepackage{amsfonts}
\usepackage{epsfig}
\usepackage{cite}
\usepackage{bbm}
\usepackage{outlines}
\usepackage{mathtools}
\usepackage[ruled,vlined]{algorithm2e}
\usepackage{caption}
\usepackage{subcaption}
\usepackage{float}
\usepackage{booktabs}
\usepackage{cancel}
\usepackage[normalem]{ulem}

\usepackage[colorlinks,bookmarksopen,bookmarksnumbered,citecolor=blue,urlcolor=red]{hyperref}
\usepackage{nomencl}
\usepackage{placeins}
\makenomenclature

\newtheorem{Pro}{Proposition}

\newtheorem{Cor}{Corollary}
\definecolor{mypink}{RGB}{220,20,130}

\definecolor{mypink}{RGB}{220,20,130}
\newcommand{\B}{\mathcal B}
\newcommand{\A}{\mathcal A}
\newcommand{\Oset}{\mathcal O}
\newcommand{\R}{\mathbb R}

\begin{document}
\title{
{
Remote~Tracking~with~State-Dependent Sensing  in Pull-Based Systems: A POMDP Framework
}
}

\author{
Jiapei Tian,
Abolfazl Zakeri,
Marian Codreanu,
and David Gundlegård%
\thanks{Jiapei Tian, Marian Codreanu, and David Gundlegård are with the Department of Science and Technology, Linköping University, 581~83 Linköping, Sweden (e-mail: \{jiapei.tian, marian.codreanu, david.gundlegard\}@liu.se).}%
\thanks{Abolfazl Zakeri is with the Centre for Wireless Communications (CWC), University of Oulu, 91004 Oulu, Finland (e-mail: abolfazl.zakeri@oulu.fi).}%
\thanks{A preliminary version~\cite{jiapei2026} is accepted to appear at IEEE WCNC 2026.}

}

\maketitle
\begin{abstract}
We consider real-time remote tracking of a Markov source observed by multiple sensors with \textit{state-dependent} sensing accuracy, motivated by applications such as industrial vehicle tracking and distributed sensing systems with heterogeneous and overlapping coverage. Upon commands from a remote sink, sensors transmit their observations over error-prone channels. We aim to minimize the long-term average of a weighted sum of goal-aware distortion and transmission costs. The problem is formulated as a partially observable Markov decision process (POMDP) and cast into an equivalent belief-MDP. To address the intractability of the infinite and continuous belief space, we develop a truncation-based method that yields a finite-state MDP which can be solved via standard methods such as relative value iteration. We further use a discounted reformulation to derive a theoretical lower bound  for the optimal average cost, which is tightened via the incremental pruning algorithm (IPA) and also induces a comparison policy. Numerical results demonstrate that the performance of the proposed policy improves with the truncation depth at the expense of computational effort, and also outperforms low-complexity baselines across a wide range of system parameters. The results also reveal a switching-type structure of the truncation-based policy over the belief simplex and quantify the impact of key system parameters, highlighting the importance of accounting for state-dependent~sensing.
\end{abstract}

\section{Introduction} \label{Sec:intro}

Real-time remote tracking is a key enabler of next-generation Internet of Things (IoT) applications and industrial automation. Examples include automated guided vehicles (AGVs) and intelligent transportation systems~\cite{Peng2019}. In these systems, edge devices observe stochastic dynamical processes and deliver status updates to remote controllers and estimators for time-critical inference and decision making.
Under stringent communication and energy constraints, the design goal is no longer to maximize delivered data volume, but to deliver the \emph{right} information \emph{at the right time} with a quality aligned with the underlying task. This shift has driven performance metrics from network-centric notions--such as throughput and packet-level delay--toward information-timeliness measures like the age of information (AoI)~\cite{kaul2012real}. 
\\\indent
While AoI captures data freshness, it is inherently content-agnostic and thus cannot differentiate updates with heterogeneous task relevance. Motivated by this limitation, distortion-based formulations have attracted increasing attention as a principled, task-aware way to quantify the discrepancy between the process of interest and its remote estimate~\cite{uysal2022semantic,pappas2021goal}. By imposing application-dependent penalties on reconstruction errors, e.g., mean squared error~\cite{huang2020real} or absolute error~\cite{salimnejad2024real}, distortion-based metrics directly embed semantic importance into policy design, guiding \emph{when, what, and how} to communicate and allocate resources in support of the underlying estimation or control objective.
\\\indent
Prior work has extensively studied distortion-oriented sampling and scheduling for remote tracking systems (e.g.,~\cite{salimnejad2024real,pappas2021goal,fountoulakis2023goal,huang2020real,yun2018optimal,sun2019sampling,ornee2026whittle,cocco2023remote,chen2024minimizing,agheli2026pull}). However, these works commonly assume that the source processes are \emph{fully observable and independently} monitored by each edge device. Such assumptions implicitly require continuous sensing and ignore the possibility of sensor-side detection failures, leading to suboptimal decisions under practical resource constraints. However, in practical distributed monitoring systems, the observability of the source can be neither perfect nor uniform across sensors. Representative applications include surveillance and target tracking using spatially distributed radar sensors~\cite{Jun2022}, cameras with overlapping fields of view~\cite{Juheon2025}, and drones for maritime search and rescue~\cite{huanran2023}. 
A similar model also applies to industrial AGV tracking~\cite{Bui2025}, where fixed infrastructure sensors monitor vehicles along predefined routes and provide heterogeneous and partially overlapping coverage. In these systems, sensing accuracy depends on the target location: detection is more reliable near the center of a sensor's coverage region and degrades near its boundaries due to reduced resolution, occlusion, or blind spots. However, this coupling between the source state and sensing reliability has received limited attention in distortion-oriented remote tracking and scheduling. We therefore consider state-dependent observability, unlike conventional models with deterministic or state-independent sensing.
\\\indent
To address the gap, we consider a remote tracking system with a Markov source being observed by multiple sensors that, upon command, transmit observations to a remote sink through unreliable channels (see Fig.~\ref{fig:sysmodel_combined}). Each sensor is characterized by \textit{state-dependent detection probabilities}, i.e.,~\textit{imperfect sensing}, reflecting realistic sensing uncertainties according to, e.g., their spatial distribution. Only one sensor can be commanded at each slot, and activating a sensor incurs a cost. Consequently, the source state is \textit{unobservable} at the sink unless a status update carrying source information is successfully received. 
Within this framework, we aim to address the problem of \textit{designing an optimal sensor selection and scheduling strategy.}
\\\indent
We consider a generic goal-aware distortion measure. Since the distortion metric depends on the source state, which is not continuously observable, the problem is formulated as a partially observable Markov decision process (POMDP). The objective is to determine the optimal sensor command strategy that minimizes the long-term average of a weighted sum of distortion and transmission costs. We then cast the POMDP into a belief-MDP with a \textit{continuous} (and inherently infinite) state space, which renders existing tools for finite-state MDP problems simply inapplicable.
To address this challenge, we propose a belief truncation method that transforms the original continuous belief-MDP into a finite-state MDP, which is then optimally solved via the relative value iteration. Specifically, we
exploit the predictable belief evolution under unsuccessful observations to truncate the belief space. Notably, this truncation yields an asymptotically optimal policy for the original POMDP problem. We further derive a theoretical lower bound on the optimal average cost through a discounted reformulation of the belief-MDP. To tighten it, we solve the discounted problem using the incremental pruning algorithm (IPA)~\cite{cassandra2013incremental}. This produces a computable lower bound certificate and an IPA-based policy for comparison. The certificate becomes increasingly tight as the discount factor approaches one.

Finally, numerical results demonstrate that the proposed truncation-based policy consistently outperforms baseline policies across a wide range of system parameters. Moreover, its performance closely matches that of the policy obtained from the discounted reformulation and remains close to the derived lower bound. These results confirm that the belief-truncation framework provides a computationally efficient and asymptotically optimal approach for the infinite-horizon average-cost problem.

\subsection{Contributions}
The main contributions of this paper are summarized as follows:
\begin{itemize}
    \item We formulate a real-time tracking problem of a partially observable source subject to \textit{state-dependent} sensing characteristics; this setting is motivated by infrastructure-assisted AGV tracking with spatially distributed sensors and partially overlapping coverage. We model the problem as a POMDP to minimize the long-term average of a weighted sum of distortion and transmission costs, explicitly accounting for the coupling between the source state, detection probability, and the distortion metric.
    \item We cast the POMDP into a continuous-state belief-MDP. To address the intractability of the infinite belief space, we develop a belief-state truncation method that yields a finite-state MDP and establishes the asymptotic optimality of the truncation-based policy. We further use a discounted reformulation to derive a theoretical lower bound on the optimal average cost. The discounted problem is solved via IPA, which computes the lower bound and also provides an IPA-based policy for comparison.
    \item We evaluate the proposed truncation-based policy against the IPA-based policy, the derived lower bound, and two low-complexity baseline policies. The results confirm the efficacy of the belief-truncation strategy in striking a balance between performance and computational complexity, and reveal a switching-type structure of the resulting scheduling policy over the belief space.

\end{itemize}
\subsection{Notations and Paper Organization}
Notations: Random variables are denoted by uppercase letters (e.g., $X$, $O$, $A$) while their realizations are denoted by lowercase letters (e.g., $x$, $o$, $a$). Vectors and matrices are represented by lowercase and uppercase boldface letters, respectively (e.g., $\mathbf{a}$ and $\mathbf{A}$). The $i$-th element of a vector $\mathbf{a}$ is denoted by $[\mathbf{a}]_i$. For a matrix $\mathbf{A}$, $[\mathbf{A}]_{i,j}$ denotes the entry in the $i$-th row and $j$-th column, while $[\mathbf{A}]_{j}$ refers to its $j$-th column, and $\mathbf{A}\trans$ denotes its transpose. Calligraphic letters (e.g., $\mathcal{X}, \mathcal{A}$) denote sets, and $|\mathcal{A}|$ denotes the cardinality of $\mathcal{A}$.  We use $\Pr\{\cdot\}$ for the probability of an event and $p(\cdot)$ for the corresponding (conditional) probability mass function (PMF), i.e., ${ p(o \mid x, a) = \Pr\{O=o \mid X=x, A=a\} }$. Similarly, we denote the probability density function (PDF) by $f(\cdot)$. The expectation operator is denoted by $\mathbb{E}\{\cdot\}$.  The indicator function is denoted by $\mathds{1}_{\{\cdot\}}$, which takes the value $1$ if the condition in the braces is satisfied and $0$ otherwise. 

Paper Organization: The remainder of the paper is organized as follows. Related work is discussed in Section~\ref{sec:related_work}. The system model and problem formulation are presented in Section~\ref{Sec_2}. Section~\ref{Sec_3} provides solutions to the modeled POMDP. The numerical results are shown in Section~\ref{Sec:Num_Re}. The paper is concluded in Section~\ref{Sec:conclusion}.

\section{Related Work}\label{sec:related_work}
Early works on real-time remote tracking established the AoI as a fundamental metric for optimizing the information freshness~\cite{kadota2018scheduling,moltafet2020age,zakeri2023minimizing}. While effective for freshness-sensitive applications, AoI remains content-agnostic, failing to account for the semantic importance of the transmitted data (status updating). To address this, recent studies have shifted towards goal-oriented performance metrics~\cite{zakeri_wcnc,salimnejad2024real,huang2020real,pappas2021goal,fountoulakis2023goal,yun2018optimal,sun2019sampling,zheng2020urgency,ornee2026whittle,chen2024minimizing,chen2025preempting,cocco2023remote,agheli2026pull,chiariotti2025goal,kam2020age,chiariotti2022scheduling,agheli2025integrated,luo2025semantic,zakeri2024semantic}. These include age of incorrect information (AoII), defined as a composite of distortion and age penalties, which captures the cumulative cost of maintaining an incorrect estimate over time. Unlike AoI or AoII, general distortion function (such as MSE) align directly with the system's operational goals, explicitly penalizing the goal-oriented distance between the information source and its estimate at the monitor. In~\cite{pappas2021goal}, the cost of actuation error (CAE) was formulated to prioritize information flow according to the specific control risks associated with maintaining incorrect state estimates. To capture the varying importance of updates, the Urgency of Information~\cite{zheng2020urgency} was proposed as a weighted distortion metric that leverages adaptive weights for context awareness. Going beyond generic MSE, \cite{chiariotti2022scheduling} extended the value of information to arbitrary functions of state, presenting a myopic~(one-step)~optimal scheduling strategy. 

In \cite{sun2019sampling}, remote estimation of a Wiener process with MSE as performance metric was studied, and it was shown that minimizing MSE is equivalent to minimizing AoI if sampling policies are state-independent. The authors of~\cite{huang2020real} investigated a hybrid automatic repeat
request (HARQ) based real-time estimation framework, deriving an optimal scheduling strategy that minimizes the long-term MSE. Similarly, the authors of~\cite{ornee2026whittle} proposed a Whittle index policy targeting the remote estimation of Gauss-Markov sources under the same MSE objective. The work~\cite{fountoulakis2023goal} investigated the real-time tracking problem of Markov sources under resource constraints, formulating the problem as a constrained MDP to minimize the CAE. To overcome the complexity of optimal solutions, they also derived a tractable sub-optimal policy using the drift-plus-penalty method. Reference~\cite{cocco2023remote} characterized the monitoring performance using state estimation entropy, revealing that reactive strategies minimizing estimation uncertainty fundamentally differ from those optimizing information freshness in random access channels.
The work~\cite{kam2020age} formulated the optimal sampling problem for symmetric binary Markov sources to minimize the AoII. By solving the associated MDP, they demonstrated that unlike standard AoI, minimizing AoII effectively aligns the scheduling policy with the goal of reducing real-time estimation errors.

The minimization of AoII under random transmission delays was addressed in~\cite{chen2024minimizing}, where the authors showed that for non-preemptive status update systems, the optimal policy follows a unit threshold structure under mild conditions. Building on this, the work in~\cite{chen2025preempting} incorporates packet preemption into the system model. By leveraging the policy improvement theorem, explicitly characterizes optimal policies that prioritize fresher updates by preempting stale transmissions. Expanding to distributed scenarios,~\cite{chiariotti2025goal} proposed a goal-oriented medium access protocol that leverages distributed belief processing to minimize the AoII. To capture the semantic importance of estimation errors, the authors of~\cite{luo2025semantic} proposed an average-cost CMDP framework under transmission frequency constraints to minimize the CAE, which employs an asymmetric distortion function to penalize state discrepancies based on their control impact. This concept was further refined in~\cite{luo2025cost} with the significance-aware age of consecutive error, incorporating non-linear aging penalties (e.g., exponential) to better differentiate between critical and normal estimation errors. 

Recent studies have shown increasing interest in systems with overlapping sensors that can yield common observations~\cite{gindullina2021age,chen2023analysis,zhou2020age,cao2024goal,Fidler20262D,tripathi2024optimizing,kalor2019minimizing,kalor2022timely}. Multiple-sensor status monitoring same process over independent communication links was studied in~\cite{gindullina2021age,chen2023analysis}. In~\cite{gindullina2021age}, the authors considered an energy-harvesting setting and characterized the resulting performance trade-offs, whereas~\cite{chen2023analysis} investigated a dual-updating system. In contrast, the work~\cite{zhou2020age} analyzed a multi-sensor system requiring an update from each sensor for source reconstruction. In~\cite{cao2024goal}, the authors extended multi-sensor paradigm to heterogeneous physical processes (e.g., plant dynamics versus environmental changes) within a networked control system. The work in~\cite{Fidler20262D} introduced the concept of 2D-AoI to quantify the freshness of information in distributed sensor networks monitoring spatio-temporal processes. The work in~\cite{tripathi2024optimizing} investigated a coupled-source model where correlated information from multiple sources is delivered to the monitor through a probabilistic protocol. 

However, the assumption of perfect source state knowledge in most of discussed works ignores the sensing limitations of practical systems, e.g.,~\cite{maatouk2020age,salimnejad2024real,pappas2021goal,fountoulakis2023goal,cocco2023remote,kam2020age,chen2024minimizing,agheli2026pull}, which are often characterized by noisy, costly, or incomplete observations. In~\cite{kalor2019minimizing,kalor2022timely}, the authors assumed a related overlapping sensor model with non-deterministic model, where sensors have state-dependent detection probability. In contrast to standard AoI-centric approaches adopted in most works above, we employ a general distortion framework that looks beyond freshness to characterize the goal-oriented importance of the information. The above observation limitations naturally lead to partially observable formulations. In such settings, the system state is not directly accessible and decisions must be made based on the observation history, or equivalently, on a belief (information) state obtained via Bayesian filtering. This belief-state representation converts the original POMDP into a fully observed belief-MDP, e.g.,~\cite{zakeri2024semantic,zakeri2025goal,vilni2024real,cosandal2025joint}. 


The works most closely related to ours are~\cite{kalor2022timely,zakeri2025goal}. Reference~\cite{zakeri2025goal} investigated a pull-based status updating paradigm with a distortion-minimization objective, offering valuable insights into goal-oriented querying and scheduling. However, the sensing model in~\cite{zakeri2025goal} did not capture state-dependent detection probability, which is a critical feature of many practical observation processes and is a distinct novelty of our study. Conversely, the work in~\cite{kalor2022timely} explicitly incorporates state-dependent sensing but evaluates performance primarily through AoI-based freshness criteria, rather than a task-driven distortion measure. In contrast, our work jointly considers state-dependent detection probabilities and a goal-aware distortion objective. This combination fundamentally alters how the estimation problem, and specifically the belief update, must be handled in designing an optimal scheduling policy.

\section{System Model and Problem Formulation }\label{Sec_2}



\begin{figure}[t!]
    \centering
    \hspace{7ex}    \includegraphics[width=0.5\textwidth,page=2]{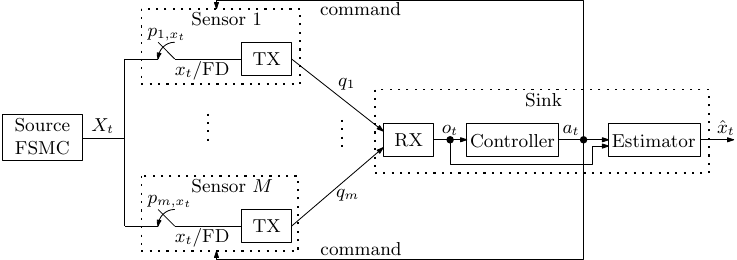}
    \caption{System model}
    \label{fig:sysmodel_combined}
\end{figure}

We consider a real-time tracking system comprising an information source, $M$ corresponding sensors and a remote sink as shown in Fig.~\ref{fig:sysmodel_combined}. We assume that time is discrete with unit time slots, i.e., ${t \in \{0,1, \dots\}}$.  The information source may encode, e.g., the (quantized) position of an object of interest such as an AGV, observed by distributed infrastructure sensors, e.g., cameras with partially overlapping fields of view~\cite{zhang2018autonomous} and position-dependent sensing reliability.
\subsubsection*{Source Model}\label{Subsec:Source_model}
We use $X_t \in \mathcal{X}$ to model the source state at slot $t$ where $\mathcal{X}=\{1,\dots,N\}$. We adopt a finite-state Markov chain (FSMC) with $N \times N$ state transition matrix $\mathbf{P}$, where its $(i,j)$ entry represents the probability of transitioning from state \( i \) to state~\( j \), i.e., \( [\mathbf{P}]_{i,j} = \Pr\{X_{t+1} = j \mid X_t = i\} \).

\subsubsection*{Imperfect Sensing Model} We assume that sensors' accuracies depend on the source state, i.e., sensor $m$ correctly senses the source state $x_t$ with probability $p_{m,x_t}$ and fails with probability ${1-p_{m,x_t}}$. Note that a commanded sensor either detects and sends the correct state information $x_t$ or, when it fails to detect the state, it sends a failed detection~(FD) notification message.
\subsubsection*{Command Actions}\label{action}
As each sensor exhibits a state-dependent performance, the sink implements an adaptive sensor selection to achieve optimal performance. We denote the command action at slot $t$ by ${a_t \in \{0,1,\ldots, M\}}$ where  $a_t=m$, $m\in \{1,\ldots, M\}$ indicates that sink commands sensor $m$ to send an update during next slot (i.e., slot $t+1$); and $a_t=0$ indicates that all sensors remain idle in slot $t+1$.

\subsubsection*{Communication Model}
To account for fading, limited transmission power, and potential interference in the wireless network, we assume that the uplink channels from sensors to the remote sink are imperfect. Let $q_m$ denote the probability that a packet transmitted by sensor $m$ is successfully received and decoded at the sink. We denote the signal decoded at the sink during slot $t$ by $o_t$. In the case of successful reception, the sink receives the sensor's message, which is either the true state $x_t$ or an FD message, i.e., $o_t \in \{x_t,\text{FD}\}$. For failed receptions (FR), $o_t=\text{FR}$. We assume error-free\footnote{Relaxing this assumption 
does not introduce fundamental difficulties, as a failed command reception simply leads to a silent sensor, which is informationally equivalent to 
a status update erasure (i.e., $\text{FR}$ observation).} channels from the sink to the sensors. 

\subsubsection*{Timing}
During slot $t$, the following events occur sequentially: (1) the sensor selected by $a_{t-1}$ (if any) transmits its status update to the sink, i.e., either true source state $x_t$ or an FD message; (2) the sink attempts to decode the received signal and generates observation $o_t\in\{x_t,\text{FD},\text{FR}\}$; (3) the controller at sink uses all currently available information to generate the command~$a_t$; and (4) the sink sends command $a_t$ to the sensors.


\subsubsection*{Performance Metric}
We quantify the discrepancy between the source state $x_t$ and its estimate at the sink $\hat{x}_t$ through a general distortion function $d \big(x_t,\hat{x}_{t}\big)$, whose form can be selected to reflect a specific goal of the application. Common choices include the error indicator $d(x,\hat x)= \mathds{1}_{\{x \neq\hat x\}}$, the absolute error $d(x,\hat x)= |x-\hat x|$, the squared error $d(x,\hat x)= (x-\hat x)^2$, or any non-negative bounded function $d:\mathcal{X} \times \mathcal{X}\rightarrow \mathbb{R}^+ $, $| d(.)|< \infty$. As an example of an application-driven design, consider a scenario where the consequences of estimation errors depend on the region to which the true state belongs. If errors within a critical region $\mathcal{R} \subset \mathcal{X}$ are more severe, we adopt an asymmetric distortion function:
\begin{equation}
\label{eq:distortion_matrix}
    d(x_t,\hat{x}_t)=
    \begin{cases}
    0 & \hat{x}_t = x_t, \\
    C_1 &   \hat{x}_t \neq x_t, x_t \in {\cal R}, \\
    C_2 &  \hat{x}_t \neq x_t, x_t \notin {\cal R}.
    \end{cases}
\end{equation}
where $C_1>C_2>0$ represent different  penalties associated with errors inside and outside the region $\mathcal{R}$.
\subsubsection*{Estimation Strategy}
The sink uses all information available at the sink until slot $t$, referred to as complete information $I_t$, and generates an estimate of the source state denoted by $\hat{x}_{t}$.\footnote{Note 
that action $a_t$ is \emph{generated locally} at the sink based on current observation $o_t$ and, potentially, on all past actions $a_0,\dots,a_{t-1}$ and all past observations $o_0,\dots,o_{t-1}$. Thus, $a_t$ provides no extra information about $X_t$ and, therefore, the complete information $I_t$ contains only \emph{past} actions $a_0,\dots,a_{t-1}$ and all observations $o_0,\dots,o_t$.} We employ a minimum distortion (MD) estimator defined as 
\begin{equation}
\label{eq:MD_est}
\hat{x}_{t}= \argmin_{x\in\mathcal{X}}~\mathbb{E}\{d(X_t,x)\mid I_t\}.    
\end{equation}
\noindent To compute the MD estimate~(\ref{eq:MD_est}), we need the distribution of the source state $X_t$ given the complete information $I_t$.   


\subsubsection*{Problem Formulation}
The objective is to determine, at each slot, the optimal command action $a_t$ that minimizes the long-term average cost, defined as a weighted sum of a distortion and a transmission cost. Formally, we state the problem as
\begin{align} \label{Equ:minimize}
      {\mbox{minimize}}~~   &
       \limsup_{T\rightarrow \infty}\,\frac{1}{T}   \sum_{t=1}^T \mathbb{E}\left\{ d(X_t, {\hat X}_t) + \alpha \mathds{1}_{\{a_t \neq 0\}} \right \},
\end{align}
with variables ${ \{a_t\}_{t=1,2,\ldots} }$, where $\alpha$ represents the cost per sensor activation\footnote{In general, different sensors may incur different activation costs, which can be modeled by assigning a sensor-dependent coefficient $\alpha_m$. To keep the formulation concise and focus on the main technical contributions, we assume identical costs across sensors, i.e., $\alpha_m=\alpha$ for all $m$.}
and the expectation is taken with respect to all sources of randomness in the system, including the source dynamics, sensing and channel impairments, and any potential randomization in the action-selection policy.


\section{Proposed solutions} \label{Sec_3}

In this section, we derive solution methods for problem~(\ref{Equ:minimize}). The main challenge arises from the fact that the sink does not have direct access to the source state, and it must instead rely on observations with state-dependent accuracy. This naturally leads to a POMDP formulation. To solve it, we first cast it into a belief-MDP and employ a truncation method to transform it into a finite-state MDP which is then optimally solved via RVIA. Furthermore, we reformulate the original belief-MDP into a discounted version and solve it using IPA. By setting the discount factor sufficiently close to one, this method provides a close-to-optimal solution. To provide practical benchmarks, we also propose two low-complexity methods.

\subsection{POMDP Formualtion}
The POMDP is specified by the following elements:
\subsubsection*{States}
The state at slot $t$ is the same as the source state $x_t$ and the state space is $\mathcal{X}$. The state is not directly observable at the sink; instead, the sink must rely only on the imperfect observations $o_t$. 
\subsubsection*{Actions}\label{Subsec:action}
The action at slot $t$ is denoted by $a_t \in\{0,\ldots,M\}$ and its meaning was discussed in Section \ref{action}. We denote the $(M+1)$-dimensional action space by $\cal A$. 
\subsubsection*{Observations} 
The observation at slot $t$ is denoted by  $o_t \in \{x_t, \text{FD}, \text{FR}\}$ and it corresponds to one of the following three cases: 
(1)~$o_t=x_t$ means that a sensor was commanded, it detected the source state $x_t$ and delivered it successfully to the sink; (2)~$o_t=\text{FD}$ (failed detection) means that a commanded sensor failed to detect the source state but successfully transmitted a failure notification message to the sink; and (3)~$o_t=\text{FR}$ (failed reception) means that the sink did not received any message during slot $t$; this can be either due to an error over the channel from sensor to sink or 
due to the fact that all sensors were idle (i.e., $a_{t-1}=0$). Formally, we define the ($N+2$)-dimensional observation space as ${\cal O} = \{1,\ldots,N,\text{FD},\text{FR}\}$.
\subsubsection*{State Transition Probabilities}
The state transitions in the POMDP follow the source dynamics and are independent of the control action, i.e., 
\begin{equation}
    p(x_{t+1} \mid x_t,a_t)=p(x_{t+1}\mid x_t),
\end{equation}
where $p(x_{t+1}\mid x_t)$ was defined in Section~\ref{Sec_2}.
\subsubsection*{Observation Function}
In general, the observation function~\cite[chapter~7.1]{sigaud2013markov} is defined as the conditional distribution of next observation 
given next state 
and current action 
, i.e., 
\begin{align}
\label{Eq:observation_function}
p(&o_{t+1} \mid x_{t+1}, a_{t}) \nonumber \\ 
& =
\begin{cases}
1 & a_{t} = 0, o_{t+1} = \text{FR}, \\
q_m p_{m,x_{t+1}} & a_{t} = m,  o_{t+1} = x_{t+1}, \\
q_m (1 - p_{m,x_{t+1}}) & a_{t} =m, o_{t+1} = \text{FD}, \\
1 - q_m & a_{t} = m ,o_{t+1} = \text{FR}, \\
0 & \text{otherwise}.
\end{cases} 
\end{align}

\subsubsection*{Cost Function}
The immediate cost at slot $t$ is defined as the sum of the distortion incurred at the sink and the transmission cost associated with the chosen action, given by 
\begin{equation}
{C_t=\E \{d(X_t,\hat{x}_t)\}+\alpha\mathds{1}_{\{a_t\neq0\}}}. 
\end{equation}

\subsection{Belief-MDP Reformulation} 
\label{seq:Belief-MDP_Formulation}
In the POMDP above, the source state is not directly accessible to the decision maker. Thus, to preserve optimality, decisions must instead be based on the entire history of past actions and observations. To avoid intractability associated with history dependence, we introduce the notion of a \emph{belief-MDP}, where the belief-states summarize all available information and, therefore, preserve Markovity~\cite[Chapter~7.3]{sigaud2013markov}.

Let $I_{t}$ denote the complete information state at slot $t$ consisting of: (1)~the initial probability distribution of the source state, (2)~all past and current observations, $o_0, \dots, o_t$, and (3)~all past actions $a_0, \dots, a_{t-1}$. We define the belief at slot $t$~as 
\begin{equation}
\mathbf{b}_t=(b_{t,1}, \ldots, b_{t,N})\trans,  
\end{equation}
where $b_{t,i}:=\text{Pr}\{X_t=i \mid I_{t}\}$. 
 \begin{Pro}\label{belief_update}
The belief $\mathbf{b}_t$ at slot $t$ is a sufficient statistic for the complete information state $I_{t}$, i.e., there exists a belief update function $\tau$ such that
\begin{equation} \label{eq:belief_update_function}
    \mathbf{b}_{t+1} = \tau\big(\mathbf{b}_{t}, a_t, o_{t+1}\big), 
\end{equation}
and belief update~(\ref{eq:belief_update_function}) 
can be expressed in terms of observation function~(\ref{Eq:observation_function}) and source transition probability matrix $\mathbf{P}$ as
\begin{equation}
\label{eq:believe_update_matrix}
 \mathbf{b}_{t+1} = k \mathbf{U}(a_t,o_{t+1})\mathbf{P}\trans\mathbf{b}_t,
\end{equation}
where $\mathbf{U}(a_t,o_{t+1})$ is a diagonal matrix given by 
\begin{align}\label{eq:diagonal_matrix}
& \mathbf{U} (a_t,o_{t+1})  \\ & \ \, = \diag \left[p( o_{t+1} \mid x_{t+1} = 1, a_t ), \ldots, p(o_{t+1} \mid x_{t+1} = N, a_t)\right] \nonumber
\end{align}
and $k$ is a normalization factor that ensures $\mathbf{1}\trans\mathbf{b}_{t+1} = 1$, given by 
\begin{equation} \label{eq:norm_k}
k = \frac{1}{\mathbf{1}\trans \mathbf{U}(a_t,o_{t+1}) \mathbf{P}\trans\mathbf{b}_t}.
\end{equation}

\end{Pro}
\begin{IEEEproof}
See Appendix \ref{seq:App_belief_sufficient}. 
\end{IEEEproof}
\begin{Cor} \label{Cor:belief_explicit}
For specific values of the action $a_t$ and the observation $o_{t+1}$, the belief update~(\ref{eq:believe_update_matrix}) can be expressed as 
\begin{align}
\label{Eq:belief_update_function}
\mathbf{b}_{t+1} = 
\begin{cases}
\mathbf{e}_n & a_t \neq 0, o_{t+1} = n,  n\in \mathcal{X}, \\
k \mathbf{U}(a_t,o_{t+1}) \mathbf{P}\trans\mathbf{b}_t & a_t \neq 0, o_{t+1} = \textnormal{FD}, \\
\mathbf{P}\trans\mathbf{b}_t & a_t\in {\cal A}, o_{t+1} = \textnormal{FR}.
\end{cases}
\end{align}
where $\mathbf{e}_n$ is $n$-th standard unit vector in $\mathbb{R}^{N}$.
\end{Cor}
\begin{IEEEproof}
It follows directly by substituting observation function $p(o_{t+1} \mid x_{t+1}, a_t)$ given by~\eqref{Eq:observation_function} into the diagonal matrix \eqref{eq:diagonal_matrix} and the resulting expression further in \eqref{eq:believe_update_matrix}.
 \end{IEEEproof}



Proposition~\ref{belief_update} ensures that the next belief $\mathbf{b}_{t+1}$ is fully determined by the current belief $\mathbf{b}_t$, the current action $a_t$, and the new observation $o_{t+1}$, without requiring the entire history of past actions and observations. Thus $\mathbf{b}_t$ can act as system state in a belief-MDP specified by the following elements:
\subsubsection*{States}
The belief-state at slot $t$ is $\mathbf{b}_t$ and the belief-state space $\mathcal{B}$ is the $(N-1)$-dimensional probability simplex, i.e., $\mathcal{B}=\{ \mathbf{b} \in\mathbb{R}^N_+ \mid  \bm{1}\trans \mathbf{b}=1 \}$. 
\subsubsection*{Actions} 
The action $a_t$ and the action space $\mathcal{A}$ are the same as in the POMDP described in~\ref{Subsec:action}. 

\subsubsection*{Belief-state transition probabilities} The transition probability from the current belief-state $\mathbf{b}_t$ to the next belief-state $\mathbf{b}_{t+1}$, given the current action $a_t$  can be expressed as:
\begin{equation} \label{eq:belief_states_density}
\begin{aligned}  
f(&\mathbf{b}_{t+1} \mid \mathbf{b}_t, a_t)  \\
& = \sum_{o_{t+1}\in\mathcal{O} } f(\mathbf{b}_{t+1},o_{t+1} \mid \mathbf{b}_t, a_t)  \\
& = \sum_{o_{t+1}\in\mathcal{O} }f(\mathbf{b}_{t+1} \mid o_{t+1},\mathbf{b}_t, a_t)p(o_{t+1}\mid \mathbf{b}_t,a_t) \\
& = \sum_{o_{t+1}\in\mathcal{O} }\delta\big(\mathbf{b}_{t+1} -\tau(\mathbf{b}_t,a_t,o_{t+1})\big) p(o_{t+1}\mid \mathbf{b}_t, a_t) ,
\end{aligned}
\end{equation}
where $\delta(\cdot)$ denotes the Dirac delta distribution and 
\begin{equation} \label{eq:next_obs_given_belief_&_action}
\begin{aligned}
 p(o_{t+1}\mid \mathbf{b}_t, a_t) 
&=  \sum_{x_{t+1}\in \mathcal{X}} p(o_{t+1}, x_{t+1} \mid \mathbf{b}_t, a_t)  \\ 
&=\sum_{x_{t+1}\in\mathcal{X}}p(o_{t+1} \mid x_{t+1}, a_t)p(x_{t+1} \mid \mathbf{b}_t)  \\
&=\sum_{x_{t+1}\in\mathcal{X}}p(o_{t+1} \mid x_{t+1},a_t) [\mathbf{P}\trans \mathbf{b}_t]_{x_{t+1}} , 
\end{aligned}
\end{equation}
where $p(o_{t+1} \mid x_{t+1},a_t)$ is the observation function~\eqref{Eq:observation_function}.


\subsubsection*{Cost Function}
Since the source state is not directly observable and the belief $\mathbf b_t$ serves as a sufficient statistic for the complete information, the immediate cost is written in its conditional expected form:
\begin{equation} \label{eq:cost_function}
\begin{aligned}  
{\overline C}_t(\mathbf{b}_t, a_t)& = \mathbb{E}_{X_t \mid \mathbf{b}_t} \{d(X_t,\hat{x}_t)\}  + \alpha \mathds{1}_{\{a_t \neq 0\}}  \\
& = \sum_{x\in \mathcal{X}} d(x, \hat{x}_t) [\mathbf{b}_t]_x + \alpha \mathds{1}_{\{a_t \neq 0\}}, 
\end{aligned}
\end{equation}
where $\hat{x}_{t}$ is the MD estimate~\eqref{eq:MD_est} of the source state 
given $\mathbf{b}_t$, 
\begin{equation}
\label{eq:MD_est_via_belief}
\hat{x}_t 
= \argmin_{x\in\mathcal{X}}~\sum_{i \in \mathcal{X}} d(i,x) [\mathbf{b}_t]_i.    
\end{equation}

The belief-MDP problem can be now cast as
\begin{align}\label{eq:belief-MDP_problem}
{\mbox{minimize}}~~  
\limsup_{T\rightarrow \infty}\,\frac{1}{T}   \sum_{t=1}^T {\overline C}_t(\mathbf{b}_t, a_t).
\end{align}

Note that the continuity of the belief state makes problem~(\ref{eq:belief-MDP_problem}) challenging, as it results in an infinite-dimensional state space $\mathcal{B}$. However, the belief update in~\eqref{Eq:belief_update_function} induces structured reachable trajectories. In the next subsection, we exploit this property by constructing a finite truncated belief set $\mathcal{B}_K$, indexed by a truncation level $K$; as $K$ increases, $\mathcal{B}_K$ contains progressively longer reachable belief trajectories and approaches the admissible reachable belief space. The policy obtained from the resulting finite-state MDP is therefore asymptotically optimal.

\subsection{Solution via Belief Space Truncation}\label{subsec:truncation}

To characterize the reachable belief-space structure, recall from Corollary~\ref{Cor:belief_explicit} that a successful source state detection and the update delivery (i.e., $o_{t+1}=n, n\in \cal{X}$) resets the belief-state to a degenerate distribution $\mathbf{b}_{t+1} = \mathbf{e}_n$. Since the source state become perfectly known at the sink, we refer to such updates as perfect. Upon receiving an imperfect update, i.e., $o_{t+1}\in\{\text{FD}, \text{FR}\}$, the belief evolves according to the Bayesian update dynamics~\eqref{Eq:belief_update_function}. For the imperfect updates, we define a \textit{belief update operator} $\mathbf{T}: \{\text{FD}, \text{FR}\}\times\cal{A} \to \mathbb{R}^{|\mathcal{X}| \times |\mathcal{X}|}$, 
\begin{equation}
    \mathbf{T}(o,a) = 
    \begin{cases} 
    \mathbf{P}\trans &  o = \mathrm{FR}, \\
    k\mathbf{U}(a,\mathrm{FD})\mathbf{P}\trans &  o = \mathrm{FD},
    \end{cases}
\end{equation}
as a matrix-valued function representing the belief update mapping for a given current observation and previous action.  
\\\indent
Starting at $t$ from a certain state $n$, i.e., $\mathbf{b}_t=\mathbf{e}_n$, consider a sequence of $K$ \textit{consecutive} imperfect observations ${o_{t+k}\in \{\text{FD}, \text{FR}\}}$ for all $k=1,\ldots,K$. 
The resulting belief state is 
\begin{equation}
    {\mathbf{b}}_{t+K} = \mathbf{T}(o_{t+K},a_{t+K-1})\cdots\mathbf{T}(o_{t+1},a_{t})\mathbf{e}_n.
\end{equation}
\indent
We denote the set of all reachable beliefs \emph{after} exactly $K$ consecutive imperfect observation by $\mathcal{R}_K$. It is given by
\begin{equation}\label{eq:reachable_belief}
\begin{aligned}
    \mathcal{R}_K = \Big\{ \big( \textstyle \prod_{k=1}^{K} \mathbf{T}(o_k,a_k) \big) \mathbf{e}_n \;\Big|\;  & o_k \in \{\text{FD}, \text{FR}\}, \\  
    & a_k \in \mathcal{A},  n \in \mathcal{X} \Big\}.
\end{aligned}
\end{equation}
Thus, the set of all reachable belief-states over \emph{any horizon} containing no more than $K$ consecutive imperfect observations is given by 
\begin{equation} \label{eq:B_K_def}
    \mathcal{B}_K = \textstyle \bigcup_{k=0}^{K} \mathcal{R}_k, 
\end{equation}
where $\mathcal{R}_0 = \{ \mathbf{e}_n \mid n\in \mathcal{X}\}$ contains all beliefs corresponding to the certain states.

The probability of $K$ consecutive imperfect observations decreases with $K$. 
Thus, the truncated set $\mathcal{B}_K$ provides an accurate approximation to the space of practically reachable belief-states for sufficiently large $K$.

Since the truncated belief-state space $\mathcal{B}_K$ is discrete, we can now define a discrete state transition probability over $\mathcal{B}_K$, as follows. For any $\mathbf{b}, \mathbf{b}'\in \mathcal{B}_K$ and $a \in \mathcal{A}$, let $p(\mathbf{b}' \mid \mathbf{b}, a)$ denote the transition probability from the current belief-state $\mathbf{b}=\mathbf{b}_t$ to the next belief-state $\mathbf{b}'=\mathbf{b}_{t+1}$, given action $a_t=a$. It can be expressed as
\begin{equation} \label{eq:truncated_state_trasition_prob}
\begin{aligned}
p(&\mathbf{b}' \mid \mathbf{b}, a) \\ 
& =
\begin{cases}
\displaystyle \sum_{o_{t+1}\in\mathcal{O} } p(\mathbf{b}',o_{t+1} \mid \mathbf{b}, a)  & \mathbf{b}'= \tau(\mathbf{b}, a, o_{t+1}), \\
0 & \text{otherwise}.
\end{cases} 
\end{aligned}
\end{equation}
where $p(\mathbf{b}',o_{t+1} \mid \mathbf{b}, a)$ is given by~\eqref{eq:belief_states_trans_prob} at the top of the next page.
\begin{figure*}[t!]
\normalsize
\begin{equation} \label{eq:belief_states_trans_prob}
p(\mathbf{b}', o_{t+1} \mid \mathbf{b} , a) =
\begin{cases}
q_{a} p_{a,n} [\mathbf{P}\trans \mathbf{b}]_n &  a\neq 0, \ \mathbf{b}'=\mathbf{e}_n, \ o_{t+1}=n, \\
q_{a} \sum_{i=1}^N (1-p_{a,i}) [\mathbf{P}\trans \mathbf{b}]_i & a\neq 0, \ \mathbf{b}'= k \mathbf{U}(a,\text{FD}) \mathbf{P}\trans\mathbf{b}, \ o_{t+1}=\text{FD}, \\
1-q_{a} &  a \neq 0, \ \mathbf{b}'= \mathbf{P}\trans\mathbf{b}, \ o_{t+1}= \text{FR}, \\
1 &  a=0, \ \mathbf{b}'= \mathbf{P}\trans\mathbf{b}, \ o_{t+1}=\text{FR} .
\end{cases}
\end{equation}
\hrulefill
\end{figure*}

In the event that the number of consecutive imperfect updates exceeds $K$, the resulting belief $\mathbf{b}'$ is replaced by its projection onto the set $\mathcal{B}_K$, i.e., the closest value contained in $\mathcal{B}_K$, given by
\begin{equation} \label{eq:projection_belief}
    \mathcal{P}_{\mathcal{B}_K}(\mathbf{b}') \in \argmin_{\mathbf{b}\in\mathcal{B}_K}~ \|\mathbf{b}'-\mathbf{b}\|_2.
\end{equation}
\indent
To account for this possibility, we replace the state transition probability~\eqref{eq:truncated_state_trasition_prob} by 
\begin{equation}
\tilde{p}(\mathbf{b}' \mid \mathbf{b}, a)
=
p(\mathbf{b}' \mid \mathbf{b}, a)
+
\sum_{\tilde{\mathbf{b}} \in \mathcal{B}_{K+1} \setminus \mathcal{B}_{K} \, : \, \mathcal{P}_{\mathcal{B}_K}(\tilde{\mathbf{b}})
=\mathbf{b}' }
p\big(\tilde{\mathbf{b}}  \mid \mathbf{b}, a\big),
\label{eq:truncated_trans_prob}
\end{equation}
where the first term represents the probability of the direct transition from $\mathbf{b}$ to $\mathbf{b}'$ and the second term accounts for the probability of the indirect transitions ${\mathbf{b} \rightarrow \tilde{\mathbf{b}} \notin \mathcal{B}_{K} \rightarrow \mathbf{b}'= \mathcal{P}_{\mathcal{B}_K}(\tilde{\mathbf{b}})}$. Specifically, the set ${\mathcal{B}_{K+1} \setminus \mathcal{B}_{K}}$ comprises all belief states outside the truncated space $\mathcal{B}_K$ that are reachable from some $\mathbf{b} \in \mathcal{B}_K$ via the update function \eqref{Eq:belief_update_function}. Within this set, the summation in \eqref{eq:truncated_trans_prob} accounts only for those beliefs that are mapped to $\mathbf{b}'$ upon projection onto $\mathcal{B}_K$. By effectively truncating the reachable belief space, we formulate a finite-state MDP solved via relative value iteration (RVIA).

\begin{Pro}\label{prop:MDPCommu}
The truncated belief-MDP is communicating.
\end{Pro}
\begin{IEEEproof}
See Appendix \ref{App_Commu}.
\end{IEEEproof}

Proposition~\ref{prop:MDPCommu} ensures the existence of a solution to the Bellman optimality equation~\cite{bertsekas2012dynamic}, i.e., there exist an optimal average cost $\rho^*$ and a relative value function $h(\mathbf{b})$ satisfying
\begin{align}
\label{eq:avg_bellman}
\rho^* + h(\mathbf{b}) = \min_{a \in \mathcal{A}} 
\Big \{
\overline{C}(\mathbf{b},a)
+
\sum_{\mathbf{b}'}
\tilde{p}(\mathbf{b}' \mid \mathbf{b}, a)\, h(\mathbf{b}')
\Big \} ,
\end{align}
for all  $\mathbf{b} \in \mathcal{B}_K$. Consequently, the actions $a^*(\mathbf{b})$ that attain the minimum in~\eqref{eq:avg_bellman} for each belief-state, define an optimal deterministic policy $\pi^*(\mathbf{b})$. 

The RVIA converts the Bellman optimality equation (\ref{eq:avg_bellman}) into an iterative procedure summarized in Algorithm \ref{alg:RVI}. By fixing an arbitrary reference state $\mathbf{b}_{\text{ref}}\in \mathcal{B}_K$ \cite[Section 4.3]{bertsekas2012dynamic}, the algorithm updates the relative value function at each iteration until convergence. Once RVIA converges, the policy 
$\pi^{*}(\mathbf{b})$ is optimal for the truncated belief-MDP~\cite[Chapter 9.4]{puterman1994markov}. 

\begin{algorithm}[t!]
\SetAlgoNoLine
\caption{Relative Value Iteration}
\label{alg:RVI}
\textbf{initialize}: $n=1$, $\epsilon$, and $V^{0}(\mathbf{b})=0$, $h^{0}(\mathbf{b}) = 0$ for all $\mathbf{b} \in \mathcal{B}_K$; select an arbitrary reference state  $\mathbf{b}_{\mathrm{ref}} \in\mathcal{B}_K$. 

\textbf{repeat} 

\ 1) for all $\mathbf{b} \in \mathcal{\mathcal{B}}_K$,
\begin{align}
V^{n}(\mathbf{b})  &=  \min_{a\in\mathcal{A}} \Big \{\overline{C}(\mathbf{b},a) + \sum_{\mathbf{b}' \in \mathcal{B}_K} \tilde{p}(\mathbf{b}' \mid \mathbf{b}, a)\, h^{n-1}(\mathbf{b}')\Big \}, \nonumber \\
h^{n}(\mathbf{b}) &= V^{n}(\mathbf{b}) - V^{n}(\mathbf{b}_{\mathrm{ref}}).  \nonumber 
\end{align} \\
\ 2) $n=n+1$ 

\vspace{.5ex}
\textbf{until} $\max_{\mathbf{b}\in \mathcal{B}_K} \big| h^{n}(\mathbf{b}) - h^{n-1}(\mathbf{b}) \big| \leq\epsilon.$

\vspace{1ex}
\textbf{optimal policy}: for all $\mathbf{b}\in \mathcal{B}_K$,
\begin{equation*}
\pi^*(\mathbf{b})
    =  \argmin_{a\in\mathcal{A}}
\Big \{ \overline{C}(\mathbf{b},a)
+
 \sum_{\mathbf{b}'\in\mathcal{B}_K}
\tilde{p}(\mathbf{b}'\mid\mathbf{b},a)\
h^{n}(\mathbf{b}')
\Big \}. \nonumber
\end{equation*} 
\end{algorithm}

\subsection{Lower Bound by Discounted Reformulation}
Following the vanishing discount method~\cite{hernandez1996average}, we study a discounted reformulation of the continuous belief-MDP to derive a computable lower bound for the original average-cost problem in~\eqref{eq:belief-MDP_problem}; this further enables us to obtain an IPA-based benchmark policy for evaluating the truncation-based solution developed in the preceding subsection. For a discount factor $\lambda \in (0,1)$, the discounted problem is given by
\begin{equation}
\label{eq:discounted_belief_problem}
{\mbox{minimize}}~~\mathbb{E}
\left\{
\sum_{t=0}^{\infty}
\lambda^t \overline{C}_t(\mathbf{b}_t,a_t)
\right\},
\end{equation}

The optimal value function associated with~\eqref{eq:discounted_belief_problem} satisfies the  following Bellman optimality equation~\cite[Theorem~7.6.1]{krishnamurthy2016pomdp}
\begin{equation}
\label{eq:bellman_discounted}
V^*(\mathbf{b})
=
\min_{a\in\mathcal A}
\Big \{
\mathbf{c}_a\trans\mathbf{b}+\lambda \sum_{o\in\mathcal O} p(o \mid \mathbf{b},a)\,V^*\big(\tau(\mathbf{b},a,o)\big)\Big \},
\end{equation}
where $p(o\mid\mathbf{b},a)$ is given by~\eqref{eq:next_obs_given_belief_&_action} and $\mathbf{c}_a$ is the immediate cost vector derived from~\eqref{eq:cost_function}, with $i$th entry ${[\mathbf{c}_a]_i =d(i,\hat{x})+\alpha \mathds{1}_{\{a \neq 0\}}}$. 

\begin{Pro}
\label{prop:span_based_lb}
For a discount factor $\lambda \in (0,1)$, define
$J_\lambda^*(\mathbf b)=(1-\lambda)V^*(\mathbf b)$ and let $J_{\mathrm{avg}}^*$ denote the optimal value of~\eqref{eq:belief-MDP_problem}. Then a lower bound is given by
\begin{equation}
\label{eq:span_based_lb}
J_{\mathrm{avg}}^* \ge
J_\lambda^*(\mathbf b)
-
(1-\lambda)\operatorname{span}(V^*(\mathbf{b}))
\end{equation}
where
\(\operatorname{span}(V^*(\mathbf b))=\sup_{\mathbf b\in\mathcal B}V^*(\mathbf b)
-\inf_{\mathbf b\in\mathcal B}V^*(\mathbf b)\).
\end{Pro}
\begin{IEEEproof}
See Appendix~\ref{seq:App_span_based_lb}.
\end{IEEEproof}
A common method to solve~\eqref{eq:bellman_discounted} is value iteration, which converts the Bellman optimality equation into an iterative procedure, i.e.,
\begin{equation}
     V^{n+1}(\mathbf{b}) = \min_{a \in \mathcal{A}} \left\{ \mathbf{c}_a\trans\mathbf{b}+\lambda \sum_{o \in \mathcal{O}} p(o \mid \mathbf{b},a)V^{n}\big(\tau(\mathbf{b},a,o)\big) \right\}.
    \label{eq:VI}
\end{equation}
Since the belief $\mathbf{b}$ lies in a continuous domain, value iteration~\eqref{eq:VI} cannot be implemented as such since it would require infinite memory. 
Thus, we use the method introduced in~\cite{cassandra2013incremental}. Specifically, we approximate the value function~\eqref{eq:VI} with a piecewise linear concave (PWLC) function parametrized by a set of supporting vectors $\mathcal{G}^n$, i.e., 
\begin{equation} \label{eq:approValue}
V^n(\mathbf b)=\min_{\bm\gamma\in\mathcal G^n}\bm\gamma\trans \mathbf b.
\end{equation}
By substituting this representation in~\eqref{eq:VI}, we obtain    
\allowdisplaybreaks
\begin{align}
&\min_{\bm{\gamma} \in \mathcal{G}^{n+1}} {\bm{\gamma}} \trans \mathbf{b} \nonumber \\ 
& = \min_{a \in \mathcal{A}} \left\{ \mathbf{c}_a\trans\mathbf{b}+\lambda \sum_{o \in \mathcal{O}}p(o \mid \mathbf{b},a) \min_{\bm{\gamma} \in \mathcal{G}^n} \bm{\gamma}\trans \tau(\mathbf{b},a,o) \right\} \nonumber \\
& = \min_{a \in \mathcal{A}} \left \{ \mathbf{c}_a\trans\mathbf{b}+\lambda \sum_{o \in \mathcal{O}}p(o \mid \mathbf{b},a) \min_{\bm{\gamma} \in \mathcal{G}^n} \bm{\gamma}\trans k\mathbf{U}(a,o)\mathbf{P}\trans\mathbf{b}  \right\}  \nonumber \\
& \stackrel{(a)}{=} \min_{a \in \mathcal{A}} \left\{ \mathbf{c}_a\trans\mathbf{b}+\lambda \sum_{o \in \mathcal{O}} \min_{\bm{\gamma}\in \mathcal{G}^n} \bm{\gamma}\trans \mathbf{U}(a,o)\mathbf{P}\trans \mathbf{b} \right \} \nonumber \\
& = \min_{a \in \mathcal{A}} \left\{\sum_{o \in \mathcal{O}} \min_{\bm{\gamma} \in \mathcal{G}^n}  \left(|\mathcal{O}|^{-1} \mathbf{c}_a+ \lambda\mathbf{P} \mathbf{U}\trans(a,o) \bm{\gamma}\right) \trans \mathbf{b} \right\} \nonumber\\
& = \min_{a \in \mathcal{A}} \left\{\sum_{o \in \mathcal{O}} \min_{\bm{\eta} \in \mathcal{G}^n_{a,o}} {\bm{\eta}} \trans \mathbf{b} \right\}, \label{eq:VI_2}
\end{align}
where 
\begin{equation} \label{eq:g^n_ao}
\mathcal{G}^n_{a,o}:=\left\{|\mathcal{O}|^{-1} \mathbf{c}_a +\lambda \mathbf{P}\mathbf{U}\trans(a,o)\bm{\gamma} \mid \bm{\gamma} \in \mathcal{G}^n\right\},
\end{equation} 
and $(a)$ follows by canceling the probability $p(o \mid \mathbf{b},a)$ with normalization factor $k$, i.e., from \eqref{eq:norm_k} we have
\begin{align}
    k & = \frac{1}{\mathbf{1}\trans \mathbf{U}(a,o) \mathbf{P}\trans\mathbf{b}}  \stackrel{(\ref{eq:diagonal_matrix})}{=} 
    \frac{1}{\displaystyle \sum^N_{x=1}p(o \mid x,a) [\mathbf{P}\trans\mathbf{b}]_{x}} 
    \stackrel{(\ref{eq:next_obs_given_belief_&_action})}{=} 
    {p(o \mid \mathbf{b},a )} . \nonumber
\end{align}
Since we have
\begin{align}
\label{eq:summin_to_crosssum_simple}
\sum_{o\in\mathcal O}\min_{\bm\eta \in\mathcal G^n_{a,o}}\bm\eta\trans\mathbf b
& =
\min_{\bm\eta_o\in\mathcal G^n_{a,o} \forall o\in\mathcal O }
\sum_{o\in\mathcal O}\bm\eta_o\trans\mathbf{b} \nonumber 
& =
\min_{\bm\gamma\in\bigoplus_{o\in\mathcal O}\mathcal G^n_{a,o}}\bm\gamma\trans\mathbf b, 
\end{align}
where $\bigoplus$ denotes the Minkowski sum, i.e., ${\mathcal{X} \bigoplus \mathcal{Y} = \{\mathbf{x}+\mathbf{y} \mid \mathbf{x} \in \mathcal{X}, \mathbf{y} \in \mathcal{Y}  \}}$, value iteration~\eqref{eq:VI_2} becomes
\begin{align}
\min_{\bm{\gamma} \in \mathcal{G}^{n+1}} {\bm{\gamma}} \trans \mathbf{b} 
& = \min_{a \in \mathcal{A}} \left\{ \min_{\bm\gamma \in \underset{o\in\mathcal O}{\bigoplus} \mathcal G^n_{a,o}}\bm\gamma \trans\mathbf b \right\} \nonumber \\
& = \min_{\bm\gamma \in \underset{a\in\mathcal A}{\bigcup} \left( \underset{o\in\mathcal O}{\bigoplus} \mathcal G^n_{a,o}\right)}\bm\gamma \trans\mathbf b,  
\end{align}
which implies that
\begin{equation}~\label{eq:union_action_set}
\mathcal G^{n+1} =  \bigcup_{a \in \mathcal A} \left( \bigoplus_{o\in\mathcal O}\mathcal G^n_{a,o}\right).
\end{equation}
\addtolength{\abovedisplayskip}{-1.5pt} 
\addtolength{\belowdisplayskip}{-1.5pt} 
The above equation converts value iteration~\eqref{eq:VI} into a set update iteration. Although the cardinality of $\mathcal{G}^{n}$ grows exponentially with the iteration index $n$,    
many vectors in $\mathcal{G}^{n}$ are redundant and can be eliminated via pruning as explained~below.
\\\indent 
A supporting vector $\overline{\bm{\gamma}} \in \mathcal{G}$ is redundant if its removal does not alter the PWLC value function, i.e., ${\min_{\bm{\gamma}\in\mathcal G}\bm\gamma\trans \mathbf b = \min_{\bm{\gamma}\in\mathcal G \setminus \{\overline{\bm{\gamma}}\}}\bm\gamma\trans \mathbf b}$. This holds if there is no belief $\mathbf{b} \in \mathcal{B}$ 
such that ${\overline{\bm{\gamma}}\trans \mathbf b < \bm{\gamma}\trans \mathbf b }$ for all $\mathbf{\bm \gamma} \in \mathcal{G} \setminus\{ \overline{\mathbf{\bm \gamma}}\}$. Equivalently, $\overline{\bm{\gamma}}$ is redundant if the optimal value $\delta^{*}$ of the following linear program is positive~\cite{Cassandra1994}: 
\begin{equation}
\begin{aligned}
\minimize_{\delta, \mathbf{b}} \quad & \delta \quad \\ \text{subject to}  \quad & \overline{\bm{\gamma}} \trans \mathbf{b}  \le  \bm{\gamma} \trans \mathbf{b} + \delta, \ \text{for all} \ \bm{\gamma}\in \mathcal{G}\setminus \{\overline{\bm{\gamma}}\}, \\
 & \mathbf{1}\trans \mathbf{b} = 1, \ \mathbf{b} \ge \mathbf{0}.
\end{aligned}
\end{equation}
The Prune$(\mathcal{G})$ procedure discards all supporting vectors $\overline{\bm \gamma} \in \mathcal{G}$ that satisfy the redundancy condition. Furthermore, as proposed in~\cite{cassandra2013incremental}, at each iteration, pruning can be performed efficiently on each set $\mathcal{G}^{n}_{a,o}$ as illustrated in Alg.~\ref{alg:incremental_pruning}.

The algorithm initializes with $\mathcal{G}^0=\{\mathbf{0}\}$ at $n=0$ and terminates when the maximum deviation between consecutive value function updates falls below a predefined threshold $\sigma$, i.e., $\max_{\mathbf{b} \in \mathcal{B}} |V^{n+1}(\mathbf{b}) - V^{n}(\mathbf{b})| < \sigma$, with $V^{n}(\mathbf{b})$ defined in \eqref{eq:approValue}. This stopping criterion can be evaluated exactly at the expense of solving an LP for each supporting vector in $\mathcal{G}^{n}$ and $\mathcal{G}^{n+1}$, or approximated by restricting the continuous maximization over $\mathbf{b} \in \mathcal{B}$ to a predefined finite grid $\hat{\mathcal{B}}$. Upon termination, the algorithm yields a final set of supporting vectors $\mathcal{G}^*$. In implementation, we replace $V^*(\mathbf{b})$ in Proposition~\ref{prop:span_based_lb} with the PWLC representation $\widehat V(\mathbf b)=\min_{\bm{\gamma}\in\mathcal G}\bm{\gamma}\trans\mathbf b$. Then $\operatorname{span}(\widehat V(\mathbf{b}))$ is computed over the belief simplex by solving the LPs. For any given $\mathbf{b} \in \mathcal{B}$, the optimal policy dictates selecting the action associated with the vector $\bm{\gamma} \in \mathcal{G}^*$ that attains the minimum in \eqref{eq:approValue}, given by $\argmin_{\bm{\gamma} \in \mathcal{G}^*} \bm{\gamma}\trans \mathbf{b}$.
\begin{algorithm}[t!]
\SetAlgoLined
\DontPrintSemicolon
\caption{Incremental Pruning}
\label{alg:incremental_pruning}
\textbf{Input} (at iteration $n$)\,: $\mathcal{G}^{n}$ \\
\textbf{Initialize}\,:
$\mathcal{S} =\emptyset$,  $\mathcal{S}_a = \{\mathbf{0}\}, \ \forall \ a\in \mathcal{A}$\;
compute $\mathcal{G}^{n}_{a,o}$ for all  $a\in \mathcal{A}$, $o\in \mathcal{O}$ using \eqref{eq:g^n_ao} \footnotemark \\ 
\For{$a=1$ \KwTo $\mathcal{|A|}$}{
    \For {$o =1 $ \KwTo $\mathcal{|O|}$}{
        $\mathcal{S}_a = \text{Prune}\left(\mathcal{S}_a \bigoplus \mathcal{G}^{n}_{a,o}\right)$\;
    }
    $\mathcal{S} = \mathcal{S} \bigcup \mathcal{S}_a$\;
}
$\mathcal{S} = \text{Prune}(\mathcal{S})$ \;
\Return $\mathcal{G}^{n+1}=\mathcal{S}$ \;
\end{algorithm}
\footnotetext{Strictly speaking, the algorithm must also store the corresponding action for each supporting vector, i.e, it must construct a set of vector-action pairs $\overline{\mathcal{G}}^{n}_{a,o} = \{(\gamma,a) \mid \gamma \in \mathcal{G}^{n}_{a,o} \}$. This is needed for generating the optimal policy after the algorithm convergence. To avoid cluttering the presentation of the algorithm, we make a slight abuse of notation and continue to write $\mathcal{G}^{n}_{a,o}$ for the set of supporting vectors, with the understanding that the corresponding action label is stored alongside each supporting vector $\bm \gamma$.}
\subsection{Low-Complexity Methods}\label{Policy:LC}
The RVIA-based policy, presented in Alg.~\ref{alg:RVI}, relies on tabular search, necessitating the enumeration of the entire state-action space, whereas the IPA-based policy requires the continuous maintenance of a potentially large set of supporting vectors. 
To mitigate these complexities, we propose two low-complexity (LC) policies in the sequel.
\subsubsection*{Cost-Agnostic LC Policy}
The cost-agnostic policy ignores transmission cost and chooses the sensor that maximizes the one-step expected probability of obtaining a successful observation, i.e., 
\begin{equation}
    a_t = \argmax_{m \in \{1,\dots,M\}}~ \sum_{i=1}^{N} q_mp_{m,i} [\mathbf{P}\trans\mathbf{b}_t]_i,
\end{equation}
where \([\mathbf{P}\trans\mathbf{b}_t]_i \) is the predicted probability of state \(i\) at slot \(t+1\), and \(p_{m,i}\) is the probability of sensor $m$ correctly sensing the source state $i$, defined in \ref{Sec_2}. 
\subsubsection*{Cost-Aware LC Policy}
The cost-agnostic policy does not consider sensing activation costs, which may lead to unnecessarily frequent transmissions. To explicitly balance distortion reduction and sensing overhead, we further propose a cost-aware policy based on a one-step look-ahead objective. Specifically, given the current belief \(\mathbf{b}_t\), the action is chosen to minimize the sum of expected next-slot distortion and current activation cost, i.e., 
\begin{equation}
\label{eq:myopic_policy}
\begin{split}
a_t = &\argmin_{a \in \mathcal{A}} \bigg\{  \alpha \mathds{1}_{\{a \neq 0\}} + \sum_{o_{t+1} \in \mathcal{O}} p(o_{t+1} \mid 
\mathbf{b}_t, a) \\
& \times \min_{\hat{x}_{t+1} \in \mathcal{X}} \mathbb{E} [d(X_{t+1}, \hat{x}_{t+1}) \mid \tau(\mathbf{b}_t, a, o_{t+1})] \bigg\},
\end{split}
\end{equation}
where $\alpha$ represents the sensor activation cost. The summation averages over all possible observations $o_{t+1}$ and the inner minimization term corresponds to the minimum distortion achievable under the posterior belief $\tau(\mathbf{b}_t, a, o_{t+1})$. This policy adaptively activates a sensor only when the expected reduction in distortion outweighs the sensing cost.

\section{Numerical Results}\label{Sec:Num_Re}
In this section, we present simulation results to evaluate the performance of the proposed policies and compare them against several baseline methods introduced in~\ref{Policy:LC}. 
\subsection{Simulation Setup}
Unless otherwise specified, the default system configuration consists of a source with $N=3$ states monitored by $M=3$ sensors; the channel reliability is set to $q_m=q=0.8$ for all $m$, and the transmission cost is set to $\alpha=0.3$. Moreover, the RVIA employs a convergence tolerance of $\epsilon=10^{-6}$ and uses the first state as the reference state, i.e., $\mathbf{b}_{\text{ref}}=\mathbf{e}_1$. For IPA, the discount factor $\lambda$ is set to 0.99, with a stopping criterion of $\sigma=10^{-6}$. 
\subsubsection*{Source Dynamics}
The source evolves as an FSMC with index-distance-dependent transitions.
For each $i,j\in\mathcal X$, we assume the self-transition probability is fixed as $[\mathbf P]_{i,i}=p$, and the remaining probability mass $1-p$ is distributed according to an exponential weighting
\begin{equation}
[\mathbf P]_{i,j}
= (1-p)\frac{\exp(-\beta |j-i|)}{\sum_{k\in\mathcal X\setminus\{i\}}\exp(-\beta |k-i|)},
\quad j\neq i,
\end{equation}
where $|j-i|$ denotes the index-distance in the discrete state space (no wrap-around).
Unless otherwise specified, we set $p=0.7$ and $\beta=1$.
\subsubsection*{Sensor Model}
\begin{figure}[!t]
    \centering
    \begin{subfigure}[b]{0.7\linewidth} 
        \centering
        \includegraphics[width=\linewidth]{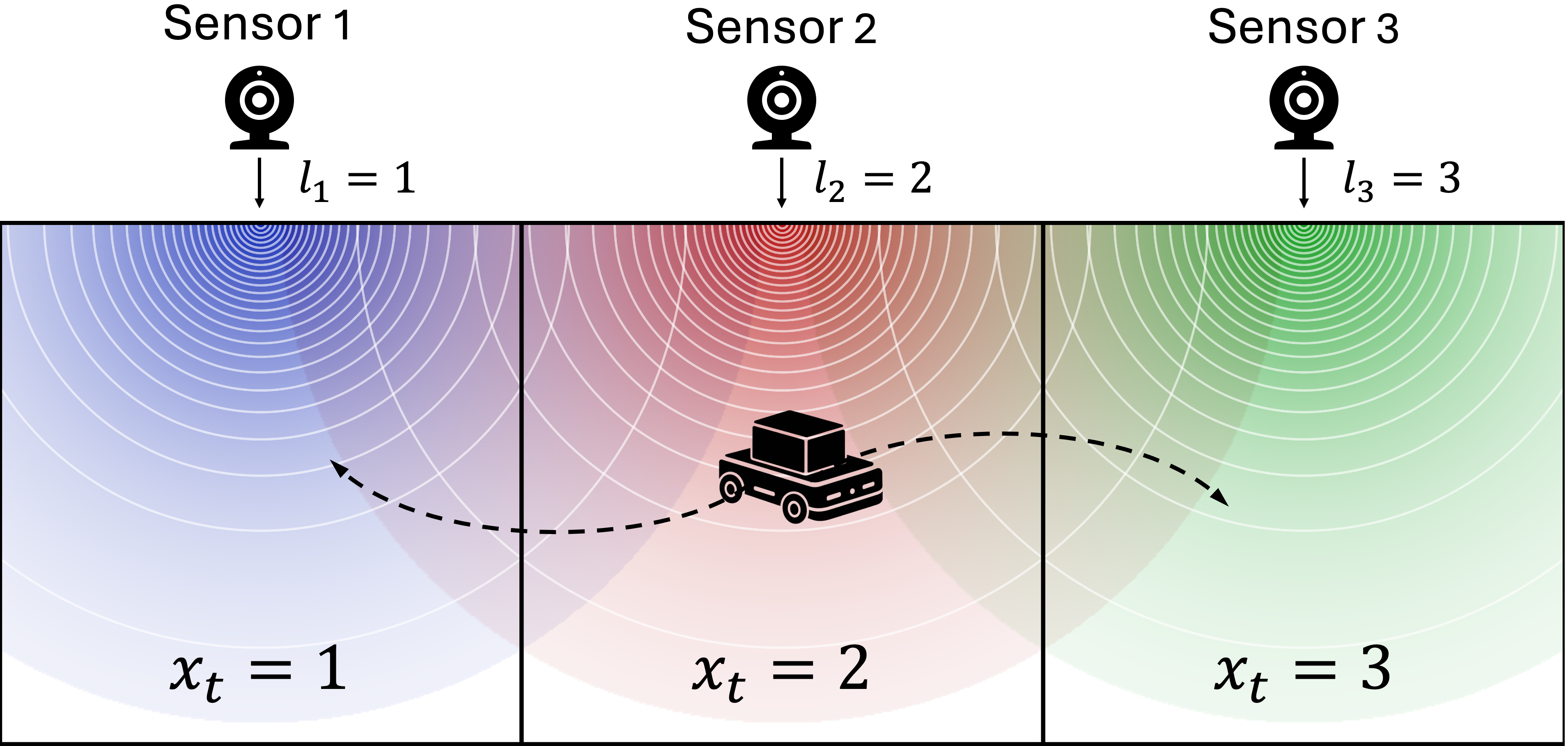}
        \caption{ $\xi=0.5$} 
        \label{fig:small_xi}
    \end{subfigure}
    
    \vspace{0.1cm} 
    
    \begin{subfigure}[b]{0.7\linewidth} 
        \centering
        \includegraphics[width=\linewidth]{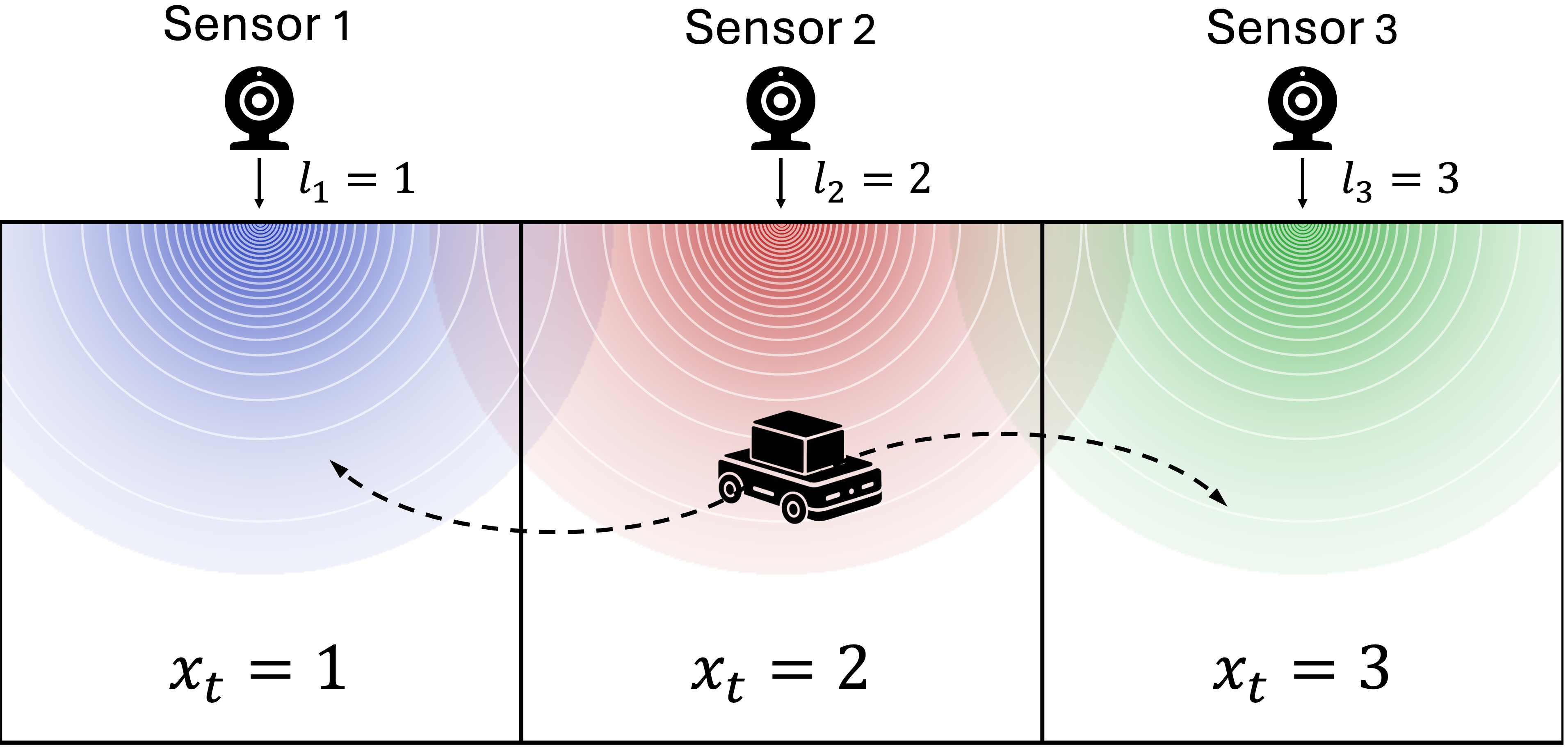}
        \caption{ $\xi=1$} 
        \label{fig:large_xi}
    \end{subfigure}
    \caption{Illustration of multi-sensor tracking scenario with overlapping area for different detection decay factors $\xi$, with a possible application to AGV monitoring in industrial environments. 
    }
    \label{fig:sensor_overlapping_combined}
\end{figure}

For the state-dependent detection probabilities introduced in Section~\ref{Sec_2}, we adopt an exponential distance-decay model.
We assume the sensor $m$ is associated with the region (state) $l_m$; under the symmetric deployment with one sensor per state considered in this section, we set $l_m=m$. The resulting geometry is illustrated in Fig.~\ref{fig:sensor_overlapping_combined}. Given $x_t=i$, the detection probability of the sensor $m$ is
\begin{equation}
p_{m,i} = p_{\max}\exp\!\big(-\xi |i - l_m|\big), \quad i\in\mathcal X,
\end{equation}
where $|i-l_m|$ denotes the index distance on the discrete state space, and no wrap-around is assumed.\footnote{Intra-state spatial variations are ignored; hence, the sensing probability is assumed uniform within each state.}
A smaller decay factor (e.g., $\xi=0.5$) yields a broader effective sensing range with substantial overlap across neighboring sensors, whereas a larger decay factor (e.g., $\xi=1$) produces a more localized sensing footprint. Unless otherwise specified, we set $p_{\max}=0.8$ and~$\xi=1$.

\subsubsection*{Distortion Metric}
For the considered three-state system, we construct an asymmetric distortion function~\eqref{eq:distortion_matrix} via the penalty matrix
\begin{equation}
\mathbf{D} =
\begin{bmatrix}
0 & 1 & 2 \\
1 & 0 & 1 \\
4 & 2 & 0
\end{bmatrix},
\label{eq:distortion}
\end{equation}
where $d(i,j) = [\mathbf D]_{i,j}$ represents the penalty incurred when the true state is $i$ and the estimate is $j$. 
\begin{figure}[!t]
    \centering
    \includegraphics[width=0.8\linewidth]{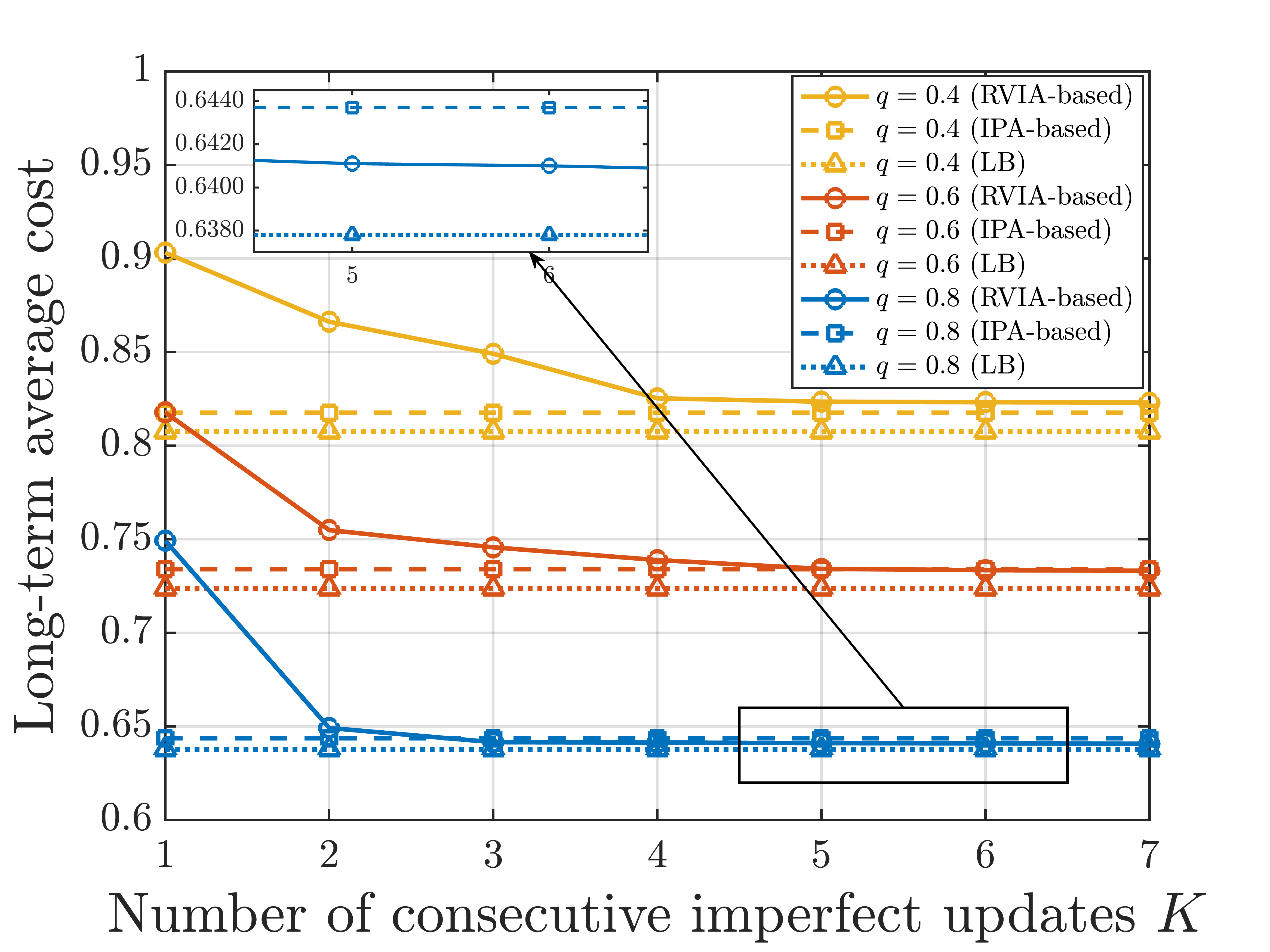}
    \caption{Long-term average cost against window size $K$ for different channel reliabilities.}
    \label{fig:performance vs K}
\end{figure}
\subsection{Performance Evaluation}
In Fig.~\ref{fig:performance vs K}, we evaluate the effectiveness of truncating the continuous belief space by examining the performance of the corresponding policies derived using the RVIA and the IPA. We also report the lower bound in Proposition~\ref{prop:span_based_lb}, computed from the IPA iterates, as a certificate for the optimal average cost. The long-term average cost is plotted against the truncation window size $K$ for different channel reliabilities $q$. As illustrated in the figure, as $q$ increases, the long-term average cost of the RVIA-based policy converges rapidly with respect to $K$. Furthermore, Fig.~\ref{fig:performance vs K} shows that beyond $K=4$, the performance improvement becomes negligible, i.e., below 1\%~improvement. Additionally, the RVIA-based policy consistently achieves a lower long-term average cost compared to the IPA-based policy for $q=0.8$ and $q=0.6$. However, we observe that for the low-reliability case of $q=0.4$, the IPA-based policy attains a slightly lower cost than the RVIA-based policy. This behavior arises because, under a highly unreliable channel, the number of consecutive unsuccessful sensing attempts increases, causing the belief state to evolve over a wider region. Consequently, a larger window size $K$ is required to obtain a good approximation of the actual belief space, and the chosen value of $K$ becomes insufficient for the RVIA-based truncation.

\begin{figure}[!t]
    \centering
    \includegraphics[width=0.8\linewidth]{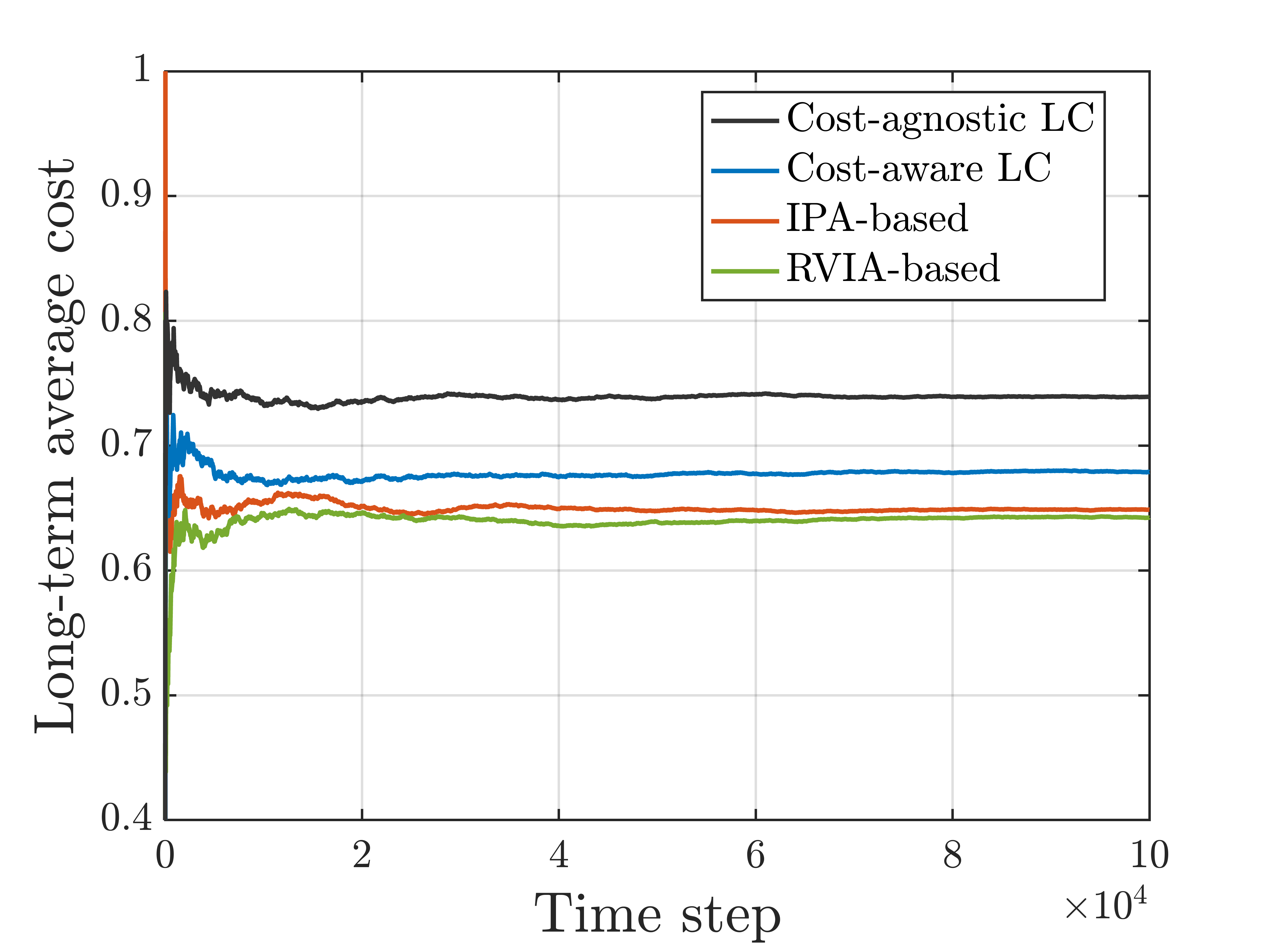}
    \caption{Long-term average cost for different policies with respect to time slots for $K=7,\ p=0.7,\ q=0.8$.}
    \label{fig:policies_performance}
\end{figure}
Fig.~\ref{fig:policies_performance} illustrates the evolution of the long-term average cost for different policies. We can observe that the RVIA-based policy consistently achieves the lowest average cost among all evaluated strategies. Although IPA-based solves optimally the discounted approximation, the truncation-based (RVIA) policy achieves a slightly lower cost.

\begin{figure}[!t]
    \centering
    \begin{subfigure}[b]{0.8\linewidth} 
        \centering
        \includegraphics[width=\linewidth]{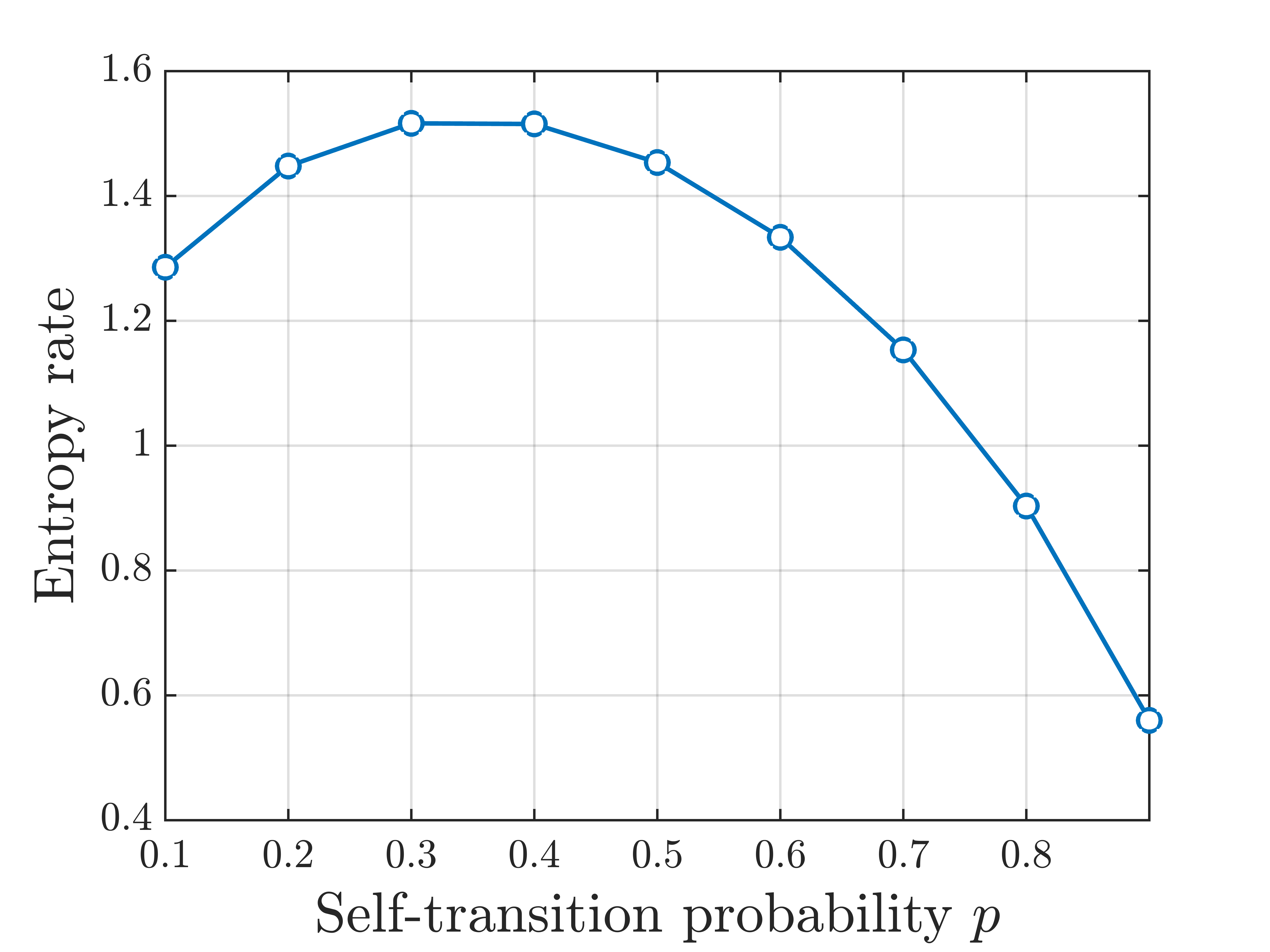}
        \caption{Source entropy rate against self-transition probability}
        \label{subfig:entrop_rate_vs_p}
    \end{subfigure}
    \medskip
    \begin{subfigure}[b]{0.8\linewidth} 
        \centering
        \includegraphics[width=\linewidth]{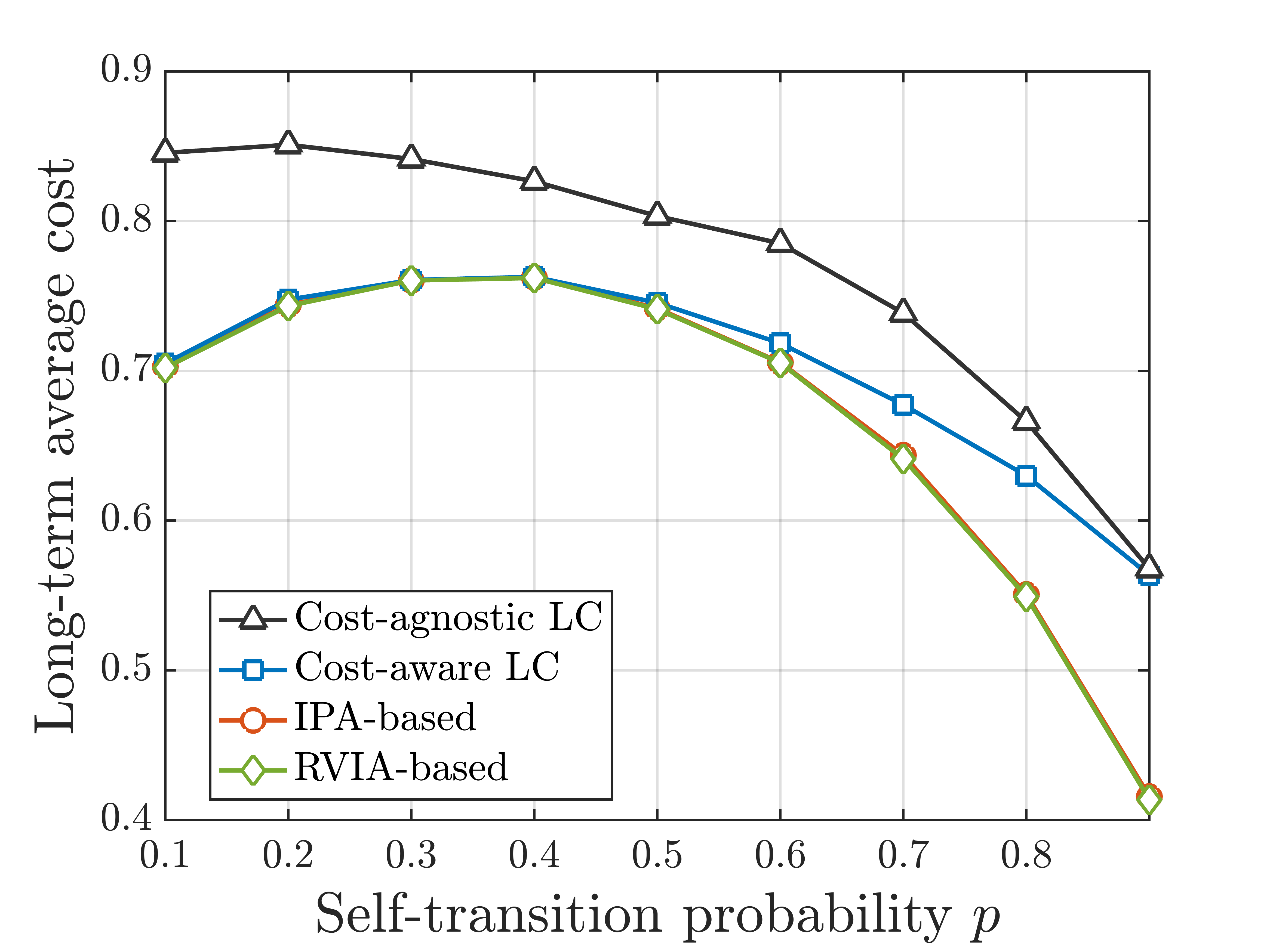}
        \caption{Long-term average cost against self-transition probability}
        \label{subfig:performance vs p}
    \end{subfigure}

    \caption{Long-term average cost and entropy rate against source self-transition probability $p$ for channel reliability $q=0.8$, transmission coefficient $\alpha=0.3$, source state transition parameters $\beta=1$.}
    \label{fig:performance vs p}
\end{figure}

Fig.~\ref{fig:performance vs p} evaluates the effect of the source dynamics, captured by the self-transition probability $p$, on the system performance. As shown in Fig.~\ref{fig:performance vs p}(\subref{subfig:entrop_rate_vs_p}), the entropy rate varies non-monotonically with $p$, indicating that source predictability does not change monotonically with the source dynamics. A key observation from Fig.~\ref{fig:performance vs p}(\subref{subfig:performance vs p}) is that the average cost largely tracks the entropy-rate trend in Fig.~\ref{fig:performance vs p}(\subref{subfig:entrop_rate_vs_p}), suggesting a tight coupling between source uncertainty and attainable performance. In Fig.~\ref{fig:performance vs p}(\subref{subfig:performance vs p}), the RVIA-based policy achieves the lowest long-term average cost over the whole range of $p$, slightly outperforming the IPA-based
, which supports the effectiveness of the truncated belief-MDP approach. The RVIA-based, IPA-based, and cost-aware LC policies reach their peak costs around $p\approx 0.4$, aligned with the maximum entropy-rate region in Fig.~\ref{fig:performance vs p}(\subref{subfig:entrop_rate_vs_p}). By contrast, the cost-agnostic LC does not explicitly balance sensing accuracy and transmission cost, and thus incurs more high-penalty errors. As $p\rightarrow 0.9$, the costs of all policies decrease, consistent with a quasi-static source, where long runs of same state make the target easier to track.

\begin{figure}[!t]
    \centering
    \includegraphics[width=0.8\linewidth]{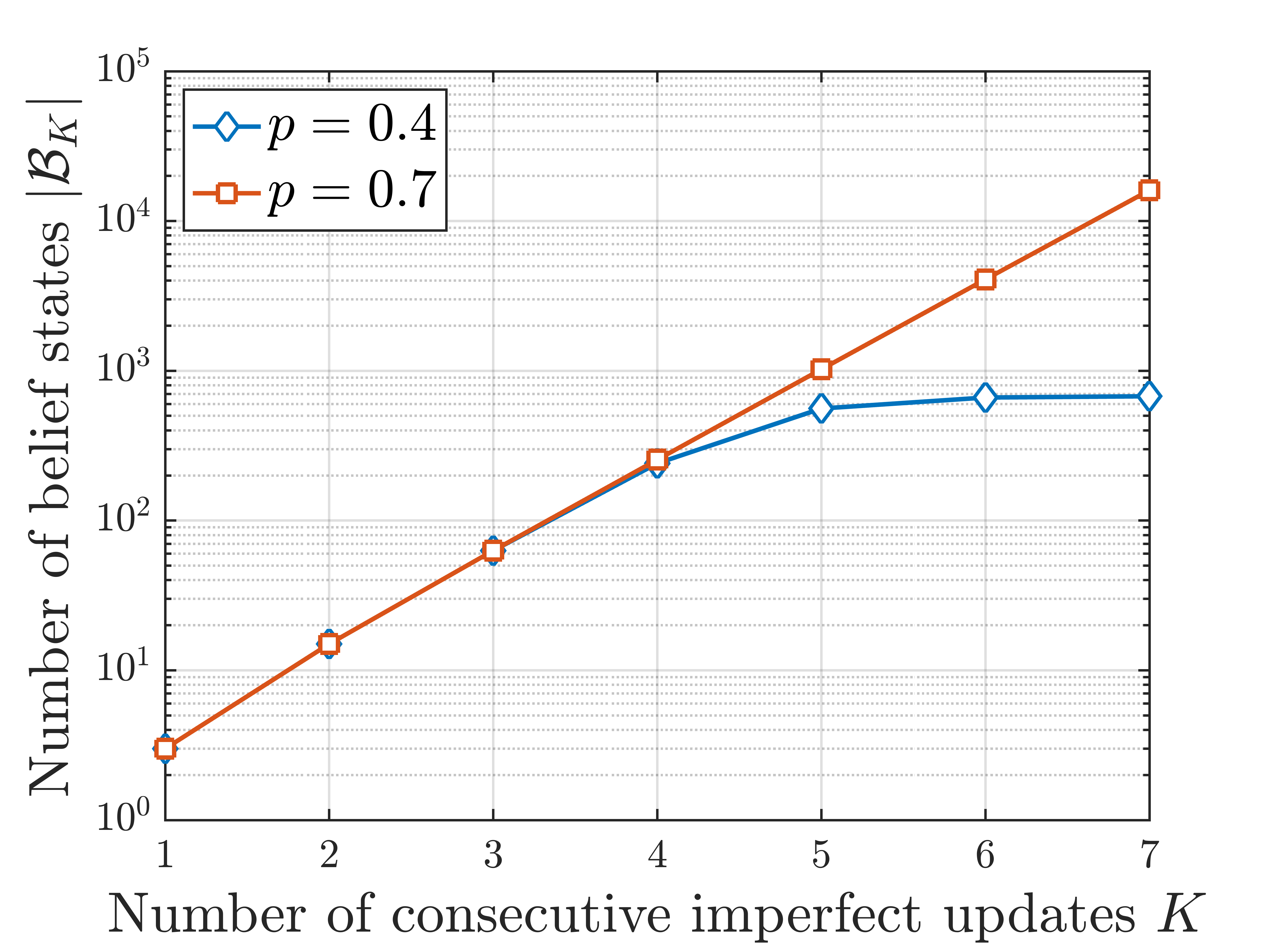}
    \caption{Cardinality of the truncated belief set $\mathcal{B}_K$ against $K$ for different source self-transition probabilities $p$ (logarithmic scale). }
    \label{fig:belief_cardinality_vs_K}
\end{figure}

Fig.~\ref{fig:belief_cardinality_vs_K} plots the number of belief states versus the truncation depth $K$ for different source self-transition probability $p$.
The vertical axis is displayed on a logarithmic scale to highlight differences in growth behavior. To provide a finite representation in our numerical construction, we merge near-identical beliefs using a $l_2$-norm tolerance of $\epsilon_m=10^{-6}$. The distinct growth behaviors are driven by the source’s temporal persistence (i.e., source dynamic).
For $p = 0.4$, the source switches more frequently and has a higher entropy rate (Fig.~\ref{fig:performance vs p}). As a result, additional history becomes redundant quickly; by $K=6$, most newly generated beliefs fall within an $\epsilon_m$-neighborhood of previously discovered beliefs, leading to early saturation.
In contrast, for $p = 0.7$ the stronger temporal persistence makes longer histories informative, so new beliefs remain separated from existing ones and $|\mathcal{B}_K|$ grows nearly exponentially with $K$ over the shown range before eventually saturating, indicating weak belief merging and a stronger dependence of computational complexity on the truncation depth.

\begin{figure}[!t]
    \centering
    \includegraphics[width=0.8\linewidth]{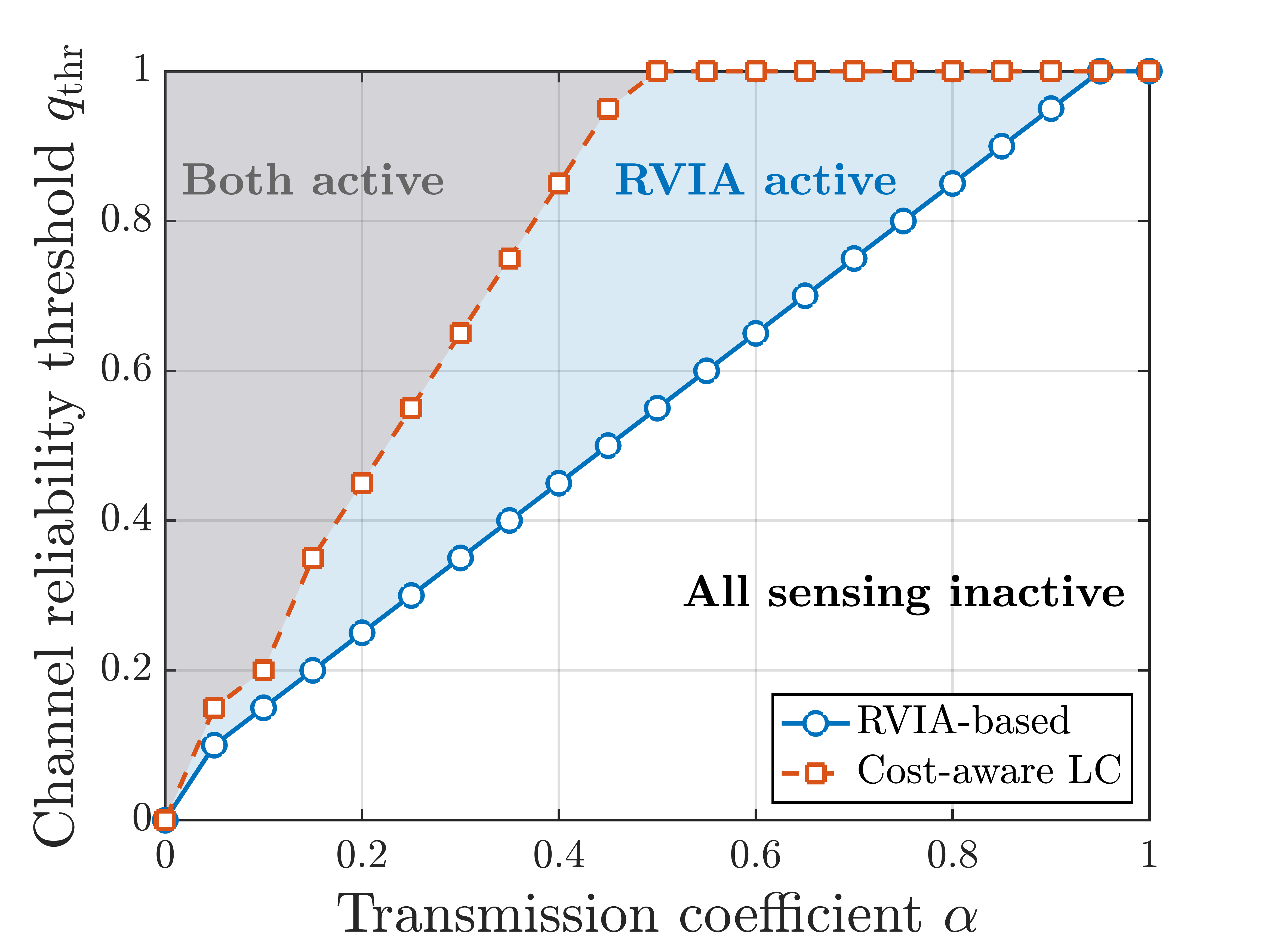}
  \caption{Channel reliability threshold $q_{\mathrm{thr}}$ against transmission coefficient $\alpha$ for the RVIA-based policy and the cost-aware LC policy. }
\label{fig:rvia_greedy_activation_regions}
\end{figure}
\begin{figure*}[!t]
    \centering
    \begin{subfigure}[t]{0.4\linewidth}
        \centering
        \includegraphics[width=\linewidth]{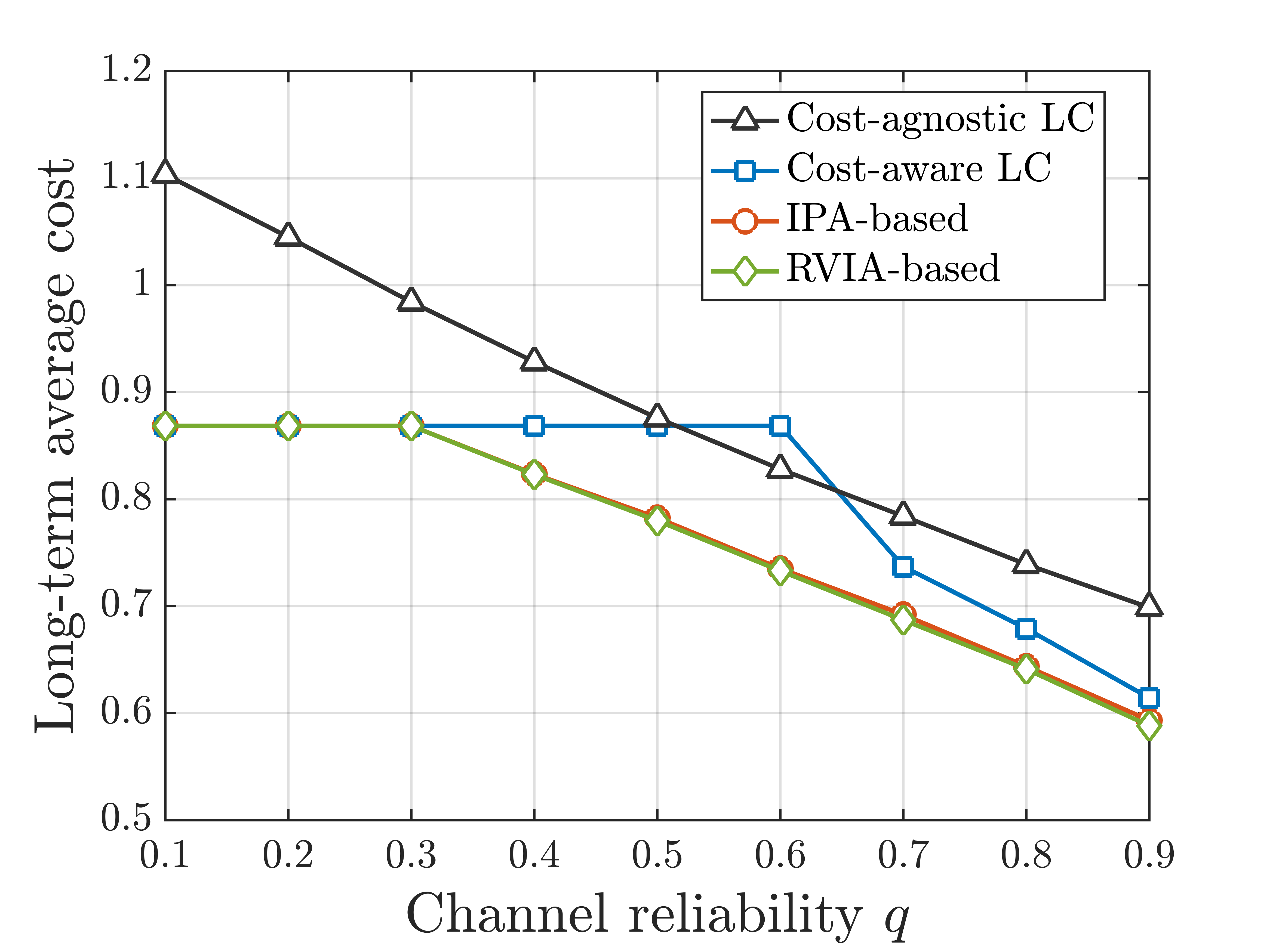}
        \caption{Impact of channel reliability $q$ with $\alpha=0.3$}
        \label{subfig:cost_vs_q}
    \end{subfigure}
     \hspace{0.2cm} 
    \begin{subfigure}[t]{0.4\linewidth}
        \centering
        \includegraphics[width=\linewidth]{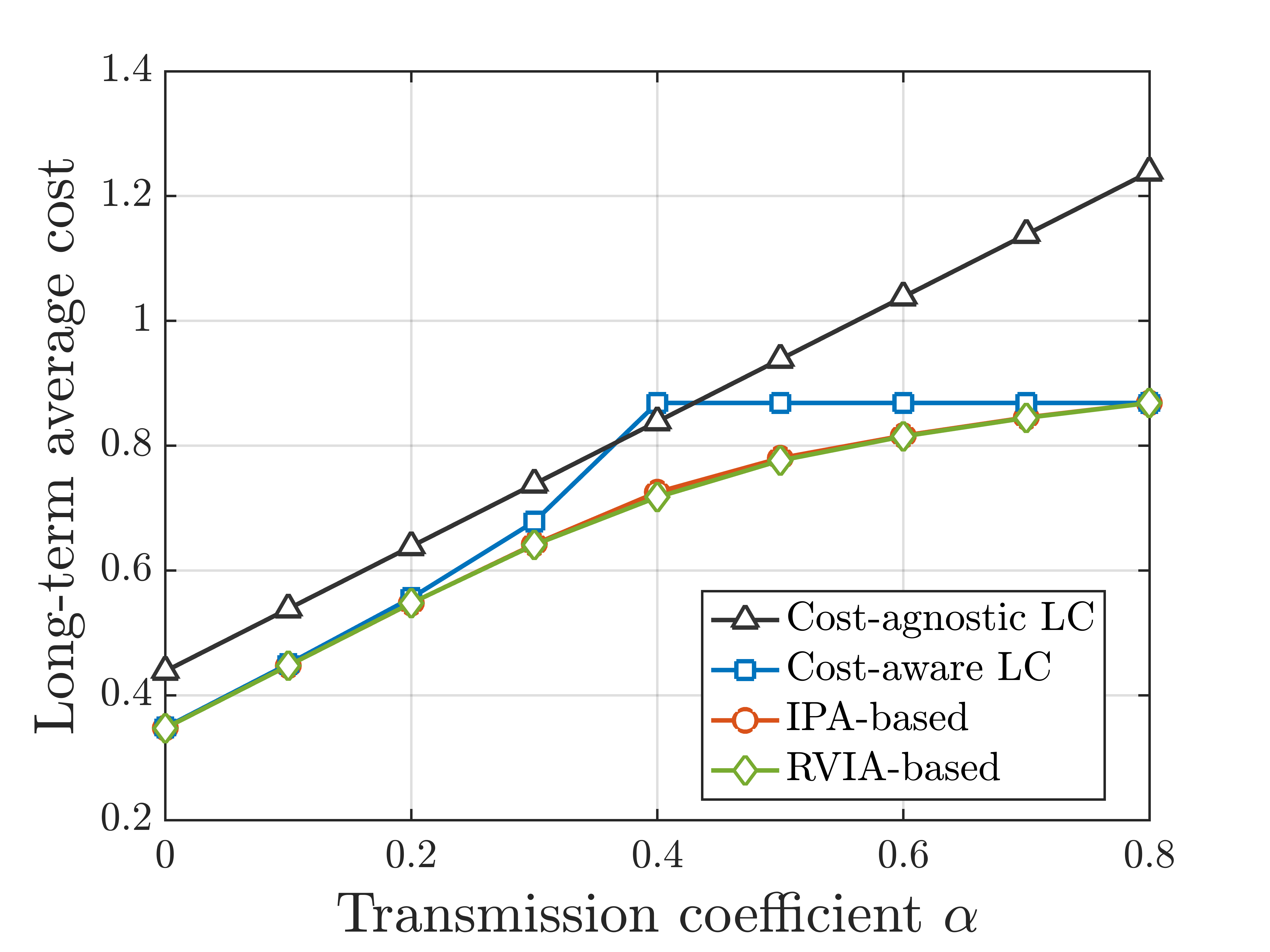}
        \caption{Impact of transmission coefficient $\alpha$}
        \label{subfig:cost_vs_alpha}
    \end{subfigure}

        \medskip
    \begin{subfigure}[t]{0.4\linewidth}
        \centering
        \includegraphics[width=\linewidth]{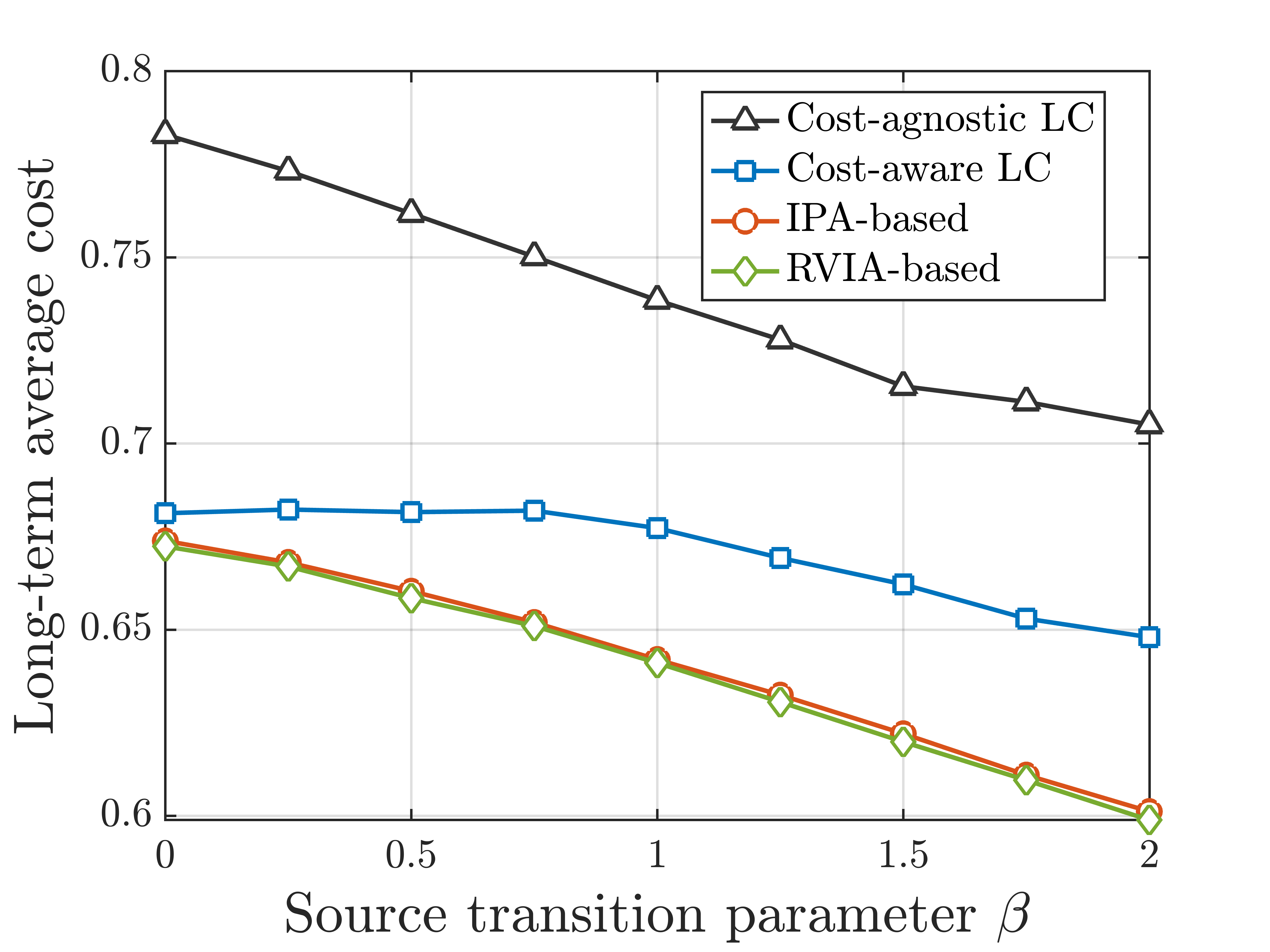}
        \caption{Impact of source transition parameter $\beta$}
        \label{subfig:cost_vs_beta}
    \end{subfigure}
    \hspace{0.2cm} 
    \begin{subfigure}[t]{0.4\linewidth}
        \centering
        \includegraphics[width=\linewidth]{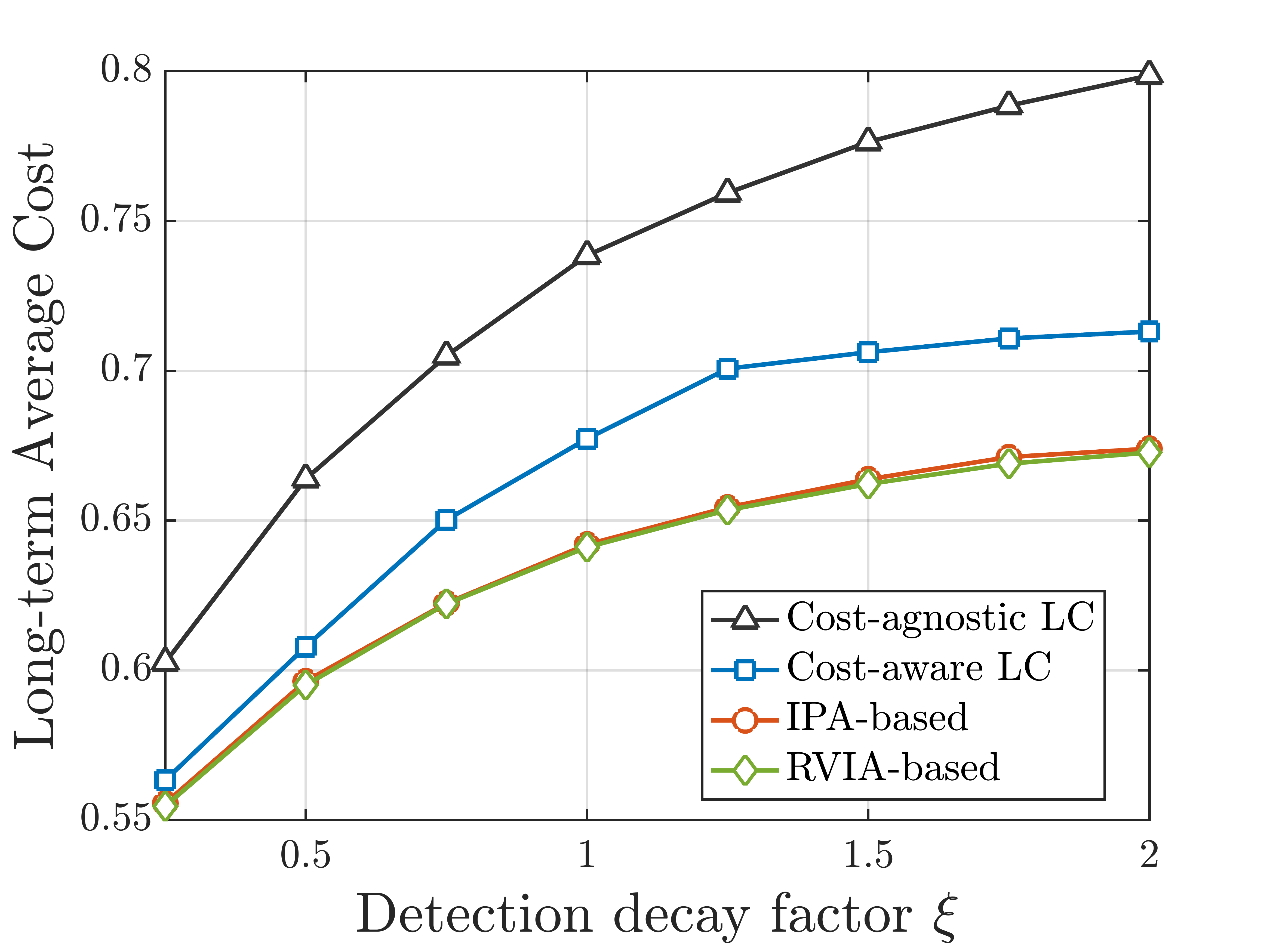}
        \caption{Impact of detection decay factor $\xi$}
        \label{subfig:cost_vs_xi}
    \end{subfigure}
    \caption{Long-term average cost versus system parameters.}
    \label{fig:combined_results}
\end{figure*}
Before investigating the impact of channel reliability $q$ and transmission cost $\alpha$ on the system performance, we first characterize the activation boundaries of the RVIA-based and cost-aware LC policy. Fig.~\ref{fig:rvia_greedy_activation_regions} determines the channel reliability threshold $q_{\mathrm{thr}}$ as a function of the transmission coefficient $\alpha$. Here, $q_{\mathrm{thr}}$ is defined as the minimum channel reliability required for a policy to schedule at least one sensor activation, i.e., scheduling a sensor is better than being idle; below this threshold is the ``\textit{All sensing inactive}'' region, where the policy deems the sensing cost prohibitive relative to the expected belief reduction and remains permanently idle. As observed, $q_{\mathrm{thr}}$ increases monotonically with $\alpha$ for both RVIA-based and cost-aware LC policies, indicating that a larger transmission penalty leads to a more conservative scheduling behavior, i.e., a higher channel quality is required to trigger sending a command for sensing and transmission. Crucially, the RVIA-based policy exhibits a consistently lower activation threshold compared to the cost-aware LC policy. This creates a distinct 
region (shaded) where the myopic cost-aware LC policy remains idle due to negative immediate returns, whereas the RVIA-based policy actively schedules transmissions. This behavior highlights the foresight of RVIA-based policy, which accepts immediate penalties under poor channel conditions to ensure long-term stability--a trade-off the cost-aware LC policy fails to capture.

Fig.~\ref{fig:combined_results} further quantifies the system performance with respect to the specific communication parameters. Fig.~\ref{fig:combined_results}(\subref{subfig:cost_vs_q}) illustrates the long-term average cost as a function of $q$. As expected, improved channel reliability reduces the cost for all policies by increasing the probability of successful status-update delivery. For low values of $q$, remaining idle is often preferable as updates rarely succeed while still incurring transmission cost. Accordingly, the cost-aware LC policy behaves conservatively and remains idle when $q$ is below its activation threshold, which explains the flat cost curve for $q < 0.6$ (see also Fig.~\ref{fig:rvia_greedy_activation_regions}). However, this policy becomes overly conservative as $q$ increases--its delayed activation around $q\approx 0.6$ causes it to miss beneficial transmissions--so it is outperformed by less conservative strategies. By contrast, the cost-agnostic LC policy schedules transmissions regardless of channel conditions; when $q$ is small, these frequent attempts largely fail and mainly accumulate transmission penalties, making it highly suboptimal. In comparison, both RVIA-based and IPA-based policies continue to achieve lower cost even in these poor channel conditions, maintaining a lower cost curve.

Similarly, Fig.~\ref{fig:combined_results}(\subref{subfig:cost_vs_alpha}) depicts the cost penalty induced by increasing the transmission coefficient $\alpha$. While the cost monotonically increases for all schemes, the cost-aware LC policy again exhibits a saturation point at high costs ($\alpha > 0.4$), where the policy stays idle for all beliefs. In contrast, both RVIA-based and IPA-based policies mitigate this penalty more effectively. By optimizing the long-term trade-off, these policies selectively schedule transmissions even at higher $\alpha$ to prevent belief divergence, thereby extending the active operational range and yielding superior cost efficiency.

Fig.~\ref{fig:combined_results} also evaluates the impact of physical system parameters. As shown in Fig.~\ref{fig:combined_results}(\subref{subfig:cost_vs_beta}), the long-term average cost decreases monotonically with the source transition parameter $\beta$. A larger $\beta$ implies that the transition probability decays rapidly with distance, confining the target's movement to local neighborhoods (i.e., high spatial correlation), which facilitates predictive tracking. Conversely, Fig.~\ref{fig:combined_results}(\subref{subfig:cost_vs_xi}) indicates that the cost rises as the detection decay factor $\xi$ increases. A higher $\xi$ shrinks the effective sensing radius and reduces the overlap between sensors (see also Fig.~\ref{fig:sensor_overlapping_combined}), thereby increasing the risk of missed detections. In both scenarios, the proposed RVIA-based and IPA-based policies yield almost identical minimum costs, significantly outperforming the baselines.
    

\begin{figure}[!t]
    \centering
    \begin{subfigure}[b]{0.48\linewidth}
        \centering
        \includegraphics[width=\linewidth]{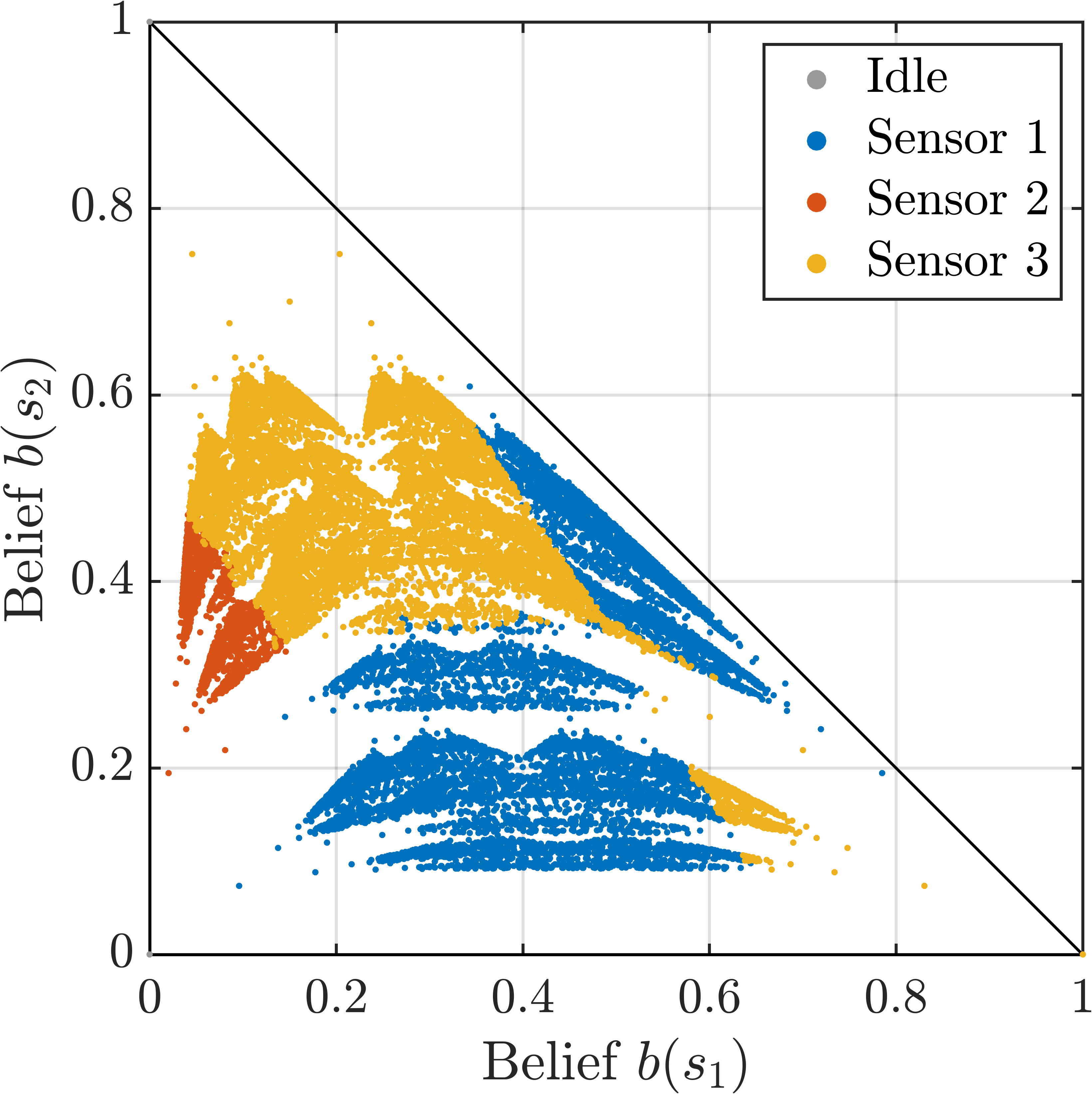}
        \caption{$\alpha=0.3$}
        \label{fig:small_cost}
    \end{subfigure}
    \hfill
    \begin{subfigure}[b]{0.48\linewidth}
        \centering
        \includegraphics[width=\linewidth]{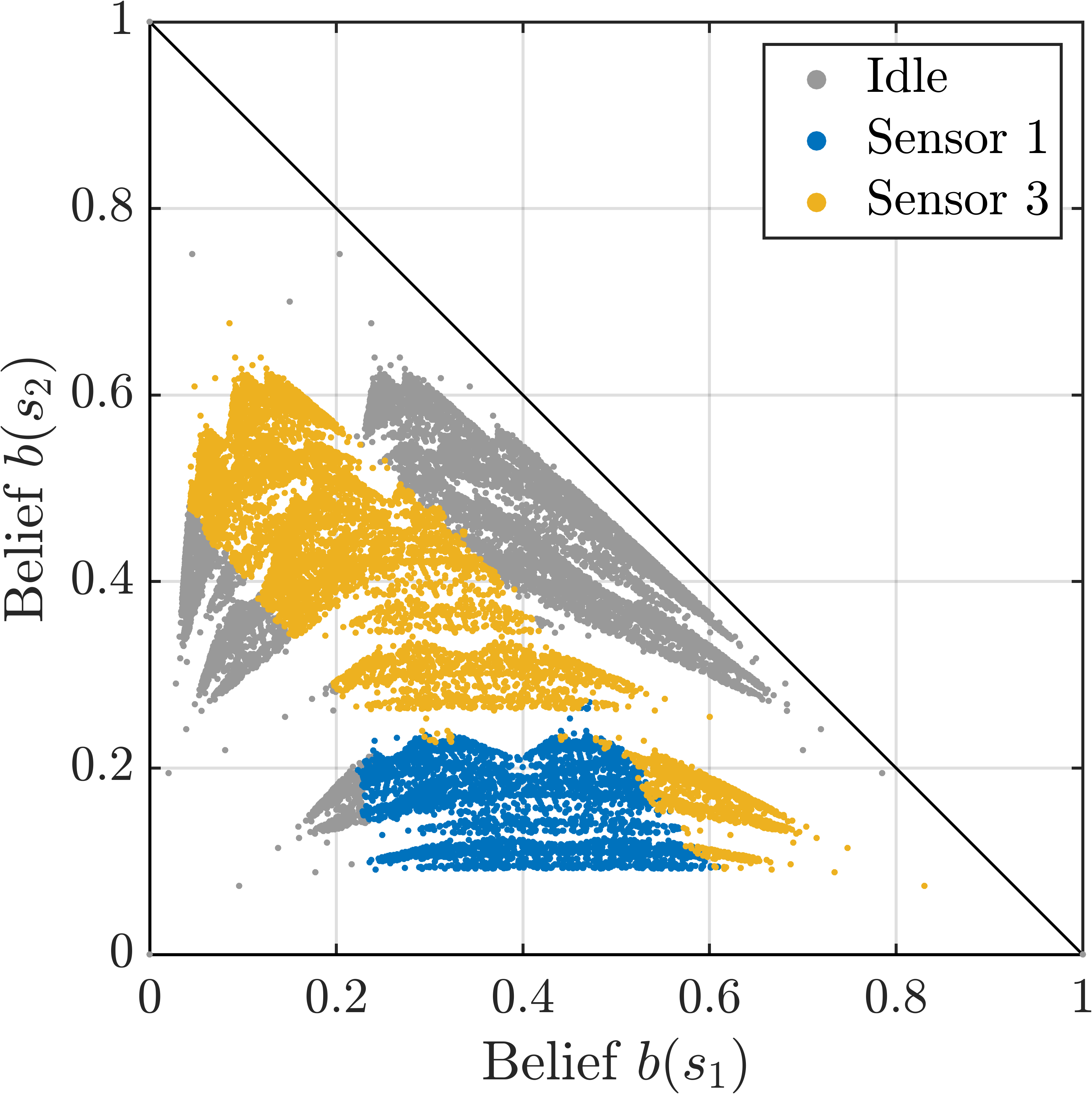}
        \caption{$\alpha=0.6$}
        \label{fig:high_cost}
    \end{subfigure}
    \vspace{-1mm}
    \caption{Structure of the RVIA-based policy with different transmission coefficients.}
    \label{fig:policy}
\end{figure}

Fig.~\ref{fig:policy} visualizes the RVIA-based policy on the belief simplex (the axes denote belief components $b(s_1)$ and $b(s_2)$, representing $\Pr\{X=1\}$ and $\Pr\{X=2\}$). Comparing the two subfigures, as the cost $\alpha$ increases, the idle region significantly expands from the vertices into the interior. These regions correspond to high-confidence belief regimes (low entropy). This implies that the optimal policy avoids unnecessary transmissions not only when the target state is perfectly known, but whenever the expected benefit of transmitting an update is outweighed by the transmission cost.

\section{CONCLUSIONS}\label{Sec:conclusion}
We investigated real-time tracking of a multi-state Markov source using heterogeneous sensors under partial observability and resource constraints. The state-dependent sensing accuracy of imperfect sensors makes the source only partially observable at the sink, while transmission costs and unreliable channels further constrain the scheduling decisions. We formulated the problem as a POMDP with the objective of minimizing the long-term average of a weighted sum of distortion and transmission costs. To address the resulting continuous belief space, we exploited the structure of belief evolution and developed a belief truncation method that yields a finite-state MDP solvable by RVIA. We also introduced a discounted reformulation to derive a lower bound reference for the optimal average cost. Since this bound depends on the span of the optimal discounted value function, we used the IPA to solve the discounted problem and also obtained an IPA-based policy for comparison.

Numerical results showed that the RVIA-based policy consistently outperforms low-complexity baseline policies across various system configurations. Moreover, its performance remains close to the derived lower bound reference and is comparable to that of the IPA-based policy, confirming the effectiveness of the proposed truncation strategy. In challenging communication regimes with low channel reliability and/or high transmission cost, the RVIA-based policy continues to schedule informative sensing actions, whereas the myopic policy tends to remain idle due to its short-horizon nature. We also analyzed the impact of physical parameters, including the source transition parameter and sensor detection parameters. Finally, visualization of the policy over the belief simplex revealed an interpretable switching-type structure, showing that the proposed policy avoids unnecessary transmissions in high-confidence belief regimes while prioritizing sensing actions in uncertain regimes.

\appendix
\allowdisplaybreaks
\subsection{Proof of Proposition \ref{belief_update}} \label{seq:App_belief_sufficient}
\begin{align}
b_{t+1, j} 
& \triangleq \Pr\{X_{t+1}=j \mid I_{t+1}\}  \nonumber \\
& \stackrel{(a)}{=} \Pr\{X_{t+1}=j \mid o_{t+1}, a_t, I_t\}  \nonumber\\ 
& = \frac{\Pr\{X_{t+1}=j, O_{t+1}=o_{t+1} \mid a_t, I_t\}}{\Pr\{O_{t+1}=o_{t+1}\mid a_t, I_t\}} \nonumber\\
& \stackrel{(b)}{=} \frac{p(o_{t+1} \mid x_{t+1}=j, a_t, \cancel{I_t}) \Pr\{X_{t+1}=j \mid a_t, I_t\}}{p(o_{t+1}  \mid a_t, I_t)} \nonumber \\
& \stackrel{(c)}{=} \frac{p(o_{t+1} \mid X_{t+1}=j, a_t)}{p(o_{t+1} \mid a_t, I_t\}} \nonumber \\
& \times \sum_{i=1}^{N} \Pr\{X_{t+1}=j \mid X_t=i, \cancel{a_t,I_t}\}
      \Pr\{X_t=i \mid \cancel{a_t},I_t\} \nonumber \\
& = \frac{p(o_{t+1} \mid x_{t+1}=j, a_t) \, [\mathbf{P}\trans \mathbf{b}_t]_j}{p(o_{t+1} \mid a_t, I_t)} \label{eq:belief_express}
\end{align}
where: (a) follows from the definition of complete information (see Section~\ref{seq:Belief-MDP_Formulation}) which implies that $I_{t+1} = \{o_{t+1}, a_t, I_t\}$; the cancellation in (b) holds true since $o_{t+1}$ is conditionally independent of $I_t$ given $\{x_{t+1}, a_t\}$; (c) follows from the law of total probability and the cancellations in (c) follow from the action-independent nature of the source transitions. 

Since the denominator in~(\ref{eq:belief_express}) is independent of $j$, it represents just a normalization factor that ensures $\mathbf{1}\trans \mathbf{b}_{t+1} = 1$, i.e., $\mathbf{b}_{t+1}$ is a legal distribution. Thus, (\ref{eq:belief_express}) is equivalent with the $j$th component of vector equality~(\ref{eq:believe_update_matrix}).  
\subsection{Proof of Proposition~\ref{prop:MDPCommu}} \label{App_Commu}
To show that the MDP is communicating, it is sufficient to find a randomized policy under which, for any ordered pair $(\mathbf b,\mathbf b')\in\mathcal{B}_K\times\mathcal{B}_K,$ the state $\mathbf b' $ is accessible from $\mathbf b$~\cite[Prop.~8.3.1]{puterman1994markov}.
Thus, the proof follows by a three-step construction: (1) reaching a certain state $\mathbf{e}_j$ from an arbitrary $\mathbf{b}$; (2) evolving the state process to arrive at the designated certain $\mathbf{e}_n$; and (3) attaining the target belief $\mathbf{b}'$ from $\mathbf{e}_n$. Consider a uniform stationary policy $\pi_{\mathrm{uni}}$ that selects each action in $\mathcal{A}$ with equal probability $\mu=1/|\mathcal{A}|>0$. The probability that $X=j$ and that $j$ is successfully detected \emph{and} delivered is $[\mathbf{b}]_{j}\sum_{a\in\mathcal{A}}\mu q_{a} p_{a,j}>0$ which reset the belief to the certain state $\mathbf{e}_j$.Secondly under $\pi_{\mathrm{uni}}$, selecting a sensor that \emph{successfully detects} state $n$ \emph{and} delivers the update occurs with positive probability (w.p.p), updating the belief $\mathbf{e}_j \to \mathbf{e}_n$. 
More precisely, since $\mathbf P$ is irreducible, there exists $L$ such that $[\mathbf P^L]_{j,n}>0$, and the event$\{X_{t+L}=n\}\cap\{\text{detect }n\text{ and deliver at }t+L\}$ has probability at least $[\mathbf P^L]_{j,n}\,\mu p_{\min}>0$ under $\pi_{\mathrm{uni}}$, where $p_{\min}:=\min_n\max_{a\in\mathcal A}q_ap_{a,n}>0$. Finally, by the definition of $\mathcal{B}_K$ in~\ref{subsec:truncation}, any $\mathbf{b}'\in\mathcal{B}_K$ is generated from the certain state $\mathbf{e}_n$ via finite consecutive imperfect observations ($k \leq K$). Since $\pi_{\mathrm{uni}}$ assigns non-zero probability to all actions, this generating sequence occurs w.p.p. 
Hence, combining the three steps above, we complete the proof.

\subsection{Proof of Proposition \ref{prop:span_based_lb}} \label{seq:App_span_based_lb}
For any bounded measurable function $f:\B\to\R$, define the belief transition operator under action $a$ and observation $o$ as
\begin{equation}
    (Q_a f)(\mathbf{b}) := \E[f(\mathbf{b}')\mid \mathbf{b},a] = \sum_{o\in\Oset}p(o\mid \mathbf{b},a)f(\tau(\mathbf{b},a,o)).
\end{equation}
The one-stage cost is denoted by $C(\mathbf{b},a)$ and under the finite-state model introduced in Sec~\ref{Sec_3}, the induced one-stage cost is bounded. Hence, there exists \(C_{\max}<\infty\) such that
$|C(\mathbf b,a)|\le C_{\max}$. For the original average-cost problem defined in~\eqref{eq:belief-MDP_problem} given a policy $\pi$, we define:
\begin{equation}
        J^\pi(\mathbf b_0):= \limsup_{T\rightarrow \infty}\,\frac{1}{T}   \sum_{t=1}^T {\overline C}_t(\mathbf{b}_t, a_t).
\end{equation}
where $\mathbf{b}_0$ denotes the initial belief,the optimal average cost is then given by
\begin{equation}
   J_{\mathrm{avg}}^*:=
    \inf_{\pi} J^\pi(\mathbf b_0).\footnote{At this point, we allow \(J_{\mathrm{avg}}^*\) to depend on the initial belief
\(\mathbf b_0\).}
\end{equation}
Define the undiscounted Bellman operator \(T\) as
\begin{equation}
    (Tf)(\mathbf b)
    :=
    \min_{a\in\A}
    \left\{
        C(\mathbf b,a)+(Q_af)(\mathbf b)
    \right\}.
\end{equation}
For any bounded \(f:\B\to\R\), define the Bellman residual
\begin{equation}
    R_f(\mathbf b):=(Tf)(\mathbf b)-f(\mathbf b),
\end{equation}
and let
\begin{equation}
    \underline r_f:=\inf_{\mathbf b\in\B}R_f(\mathbf b),
    \qquad
    \overline r_f:=\sup_{\mathbf b\in\B}R_f(\mathbf b).
\end{equation}
We first show that, for any bounded
\(f\), the residual lower and upper bounds satisfy
\begin{equation}
    \underline r_f
    \le
   J_{\mathrm{avg}}^*
    \le
    \overline r_f,
    \qquad \forall \ \mathbf b_0\in\B .
    \label{eq:residual_average_bound}
\end{equation}
Indeed, from the definition of \(T\), for every \(a\in\A\),
\begin{equation}
    C(\mathbf b,a)+(Q_af)(\mathbf b)\ge (Tf)(\mathbf b)
    \ge f(\mathbf b)+\underline r_f .
\end{equation}
Thus,
\begin{equation}
    C(\mathbf b,a)
    \ge
    \underline r_f+f(\mathbf b)-(Q_af)(\mathbf b).
\end{equation}
Applying this inequality along the trajectory induced by any admissible policy
\(\pi\), taking expectations, and using
\[
    (Q_{a_t}f)(\mathbf b_t)
    =
    \E^\pi[f(\mathbf b_{t+1})\mid \mathbf b_t,a_t],
\]
we obtain
\begin{equation}
    \E^\pi[C(\mathbf b_t,a_t)]
    \ge
    \underline r_f
    +
    \E^\pi[f(\mathbf b_t)]
    -
    \E^\pi[f(\mathbf b_{t+1})].
\end{equation}
Summing over \(t=0,\ldots,T-1\) gives
\begin{equation}
\begin{aligned}
    \E^\pi\left[
        \sum_{t=0}^{T-1} C(\mathbf{b}_t,a_t)
    \right]
    &\ge
    \sum_{t=0}^{T-1}
    \left(
        \underline r_f
        +
        \E^\pi[f(\mathbf{b}_t)]
        -
        \E^\pi[f(\mathbf{b}_{t+1})]
    \right) \\
    &=
    T\underline r_f
    +
    \sum_{t=0}^{T-1}
    \left(
        \E^\pi[f(\mathbf{b}_t)]
        -
        \E^\pi[f(\mathbf{b}_{t+1})]
    \right).
\end{aligned}
\end{equation}
The last summation is a telescoping sum:
\begin{equation}
\begin{aligned}
    \sum_{t=0}^{T-1}
    \left(
        \E^\pi[f(\mathbf{b}_t)]
        -
        \E^\pi[f(\mathbf{b}_{t+1})]
    \right)
    &=
    \E^\pi[f(\mathbf{b}_0)]
    -
    \E^\pi[f(\mathbf{b}_T)] \\
    &=
    f(\mathbf{b}_0)
    -
    \E^\pi[f(\mathbf{b}_T)],
\end{aligned}
\end{equation}
where the last equality follows because the initial belief \(\mathbf{b}_0\) is fixed. Therefore,
\begin{equation}
    \E^\pi\left[
        \sum_{t=0}^{T-1} C(\mathbf{b}_t,a_t)
    \right]
    \ge
    T\underline r_f
    +
    f(\mathbf{b}_0)
    -
    \E^\pi[f(\mathbf{b}_T)].
\end{equation}
Dividing by \(T\),
\begin{equation}
    \frac1T
    \E^\pi\left[
        \sum_{t=0}^{T-1} C(\mathbf{b}_t,a_t)
    \right]
    \ge
    \underline r_f
    +
    \frac{f(\mathbf{b}_0)-\E^\pi[f(\mathbf{b}_T)]}{T}.
\end{equation}
Since \(f\) is bounded, dividing by \(T\) and letting \(T\to\infty\) yields
\(J^\pi(\mathbf b_0)\ge \underline r_f\). Taking the infimum over all policies
gives
\begin{equation}
   J_{\mathrm{avg}}^*\ge \underline r_f .
\end{equation}

For the upper bound, let \(a_f(\mathbf b)\) be a measurable greedy selector
attaining the minimum in \(Tf\). Since \(\A\) is finite, such a selector exists.
Then
\begin{equation}
    C(\mathbf b,a_f(\mathbf b))+(Q_{a_f}f)(\mathbf b)
    =
    (Tf)(\mathbf b)
    \le
    f(\mathbf b)+\overline r_f ,
\end{equation}
which implies
\begin{equation}
    C(\mathbf b,a_f(\mathbf b))
    \le
    \overline r_f+f(\mathbf b)-(Q_{a_f}f)(\mathbf b).
\end{equation}
Repeating the same telescoping argument under the stationary policy \(a_f\)
gives
\begin{equation}
    J^{a_f}(\mathbf b_0)\le \overline r_f .
\end{equation}
Since \(J_{\mathrm{avg}}^*\le J^{a_f}(\mathbf b_0)\), we obtain
\(J_{\mathrm{avg}}^*\le\overline r_f\), proving~\eqref{eq:residual_average_bound}. Now we consider the discounted belief-MDP with discount factor \(\lambda\in(0,1)\).
Its optimal value function \(V^*\) satisfies
\begin{equation}
    V^*(\mathbf b)
    =
    \min_{a\in\A}
    \left\{
        C(\mathbf b,a)+\lambda(Q_aV^*)(\mathbf b)
    \right\}.
    \label{eq:discounted_bellman_app}
\end{equation}
Since \(C(\mathbf b,a)\) is bounded and \(\lambda<1\), \(V^*(\mathbf{b})\) is bounded. Define
\begin{equation}
    m_\lambda:=\inf_{\mathbf b\in\B}V^*(\mathbf b),
    \qquad
    M_\lambda:=\sup_{\mathbf b\in\B}V^*(\mathbf b),
\end{equation}
and
\begin{equation}
    \operatorname{span}(V^*(\mathbf{b})):=M_\lambda-m_\lambda .
\end{equation}
Let
\begin{equation}
    R_\lambda(\mathbf b):=(TV^*)(\mathbf b)-V^*(\mathbf b).
\end{equation}
We next show that
\begin{equation}
    (1-\lambda)m_\lambda
    \le
    R_\lambda(\mathbf b)
    \le
    (1-\lambda)M_\lambda,
    \qquad \forall \mathbf b\in\B .
    \label{eq:discounted_residual_bound}
\end{equation}
For the lower bound, for any \(a\in\A\),
\begin{equation}
\begin{aligned}
    C(&\mathbf b,a)+(Q_aV^*)(\mathbf b)\\ 
    &=
    C(\mathbf b,a)+\lambda(Q_aV^*)(\mathbf b)
    +(1-\lambda)(Q_aV^*)(\mathbf b)  \\
    &\ge
    C(\mathbf b,a)+\lambda(Q_aV^*)(\mathbf b)
    +(1-\lambda)m_\lambda .
\end{aligned}
\end{equation}
Taking the minimum over \(a\) and using~\eqref{eq:discounted_bellman_app}
yields
\begin{equation}
    (TV^*)(\mathbf b)\ge V^*(\mathbf b)+(1-\lambda)m_\lambda .
\end{equation}
Thus,
\begin{equation}
    R_\lambda(\mathbf b)\ge (1-\lambda)m_\lambda .
\end{equation}
For the upper bound, let
\[
    a_\lambda(\mathbf b)
    \in
    \arg\min_{a\in\A}
    \left\{
        C(\mathbf b,a)+\lambda(Q_aV^*)(\mathbf b)
    \right\}.
\]
Then
\begin{equation}
    V^*(\mathbf b)
    =
    C(\mathbf b,a_\lambda(\mathbf b))
    +
    \lambda(Q_{a_\lambda}V^*)(\mathbf b).
\end{equation}
Since \(T\) minimizes the undiscounted one-step expression,
\begin{equation}
\begin{aligned}
    (TV^*)(\mathbf b)
    &\le
    C(\mathbf b,a_\lambda(\mathbf b))
    +(Q_{a_\lambda}V^*)(\mathbf b) \\
    &=
    V^*(\mathbf b)
    +(1-\lambda)(Q_{a_\lambda}V^*)(\mathbf b) \\
    &\le
    V^*(\mathbf b)+(1-\lambda)M_\lambda .
\end{aligned}
\end{equation}
Hence,
\begin{equation}
    R_\lambda(\mathbf b)\le (1-\lambda)M_\lambda,
\end{equation}
which proves~\eqref{eq:discounted_residual_bound}.

Applying the residual bound~\eqref{eq:residual_average_bound} with
\(f=V^*\), and using the definition $R_\lambda(\mathbf b):=(TV^*)(\mathbf b)-V^*(\mathbf b)$ and~\eqref{eq:discounted_residual_bound}, gives
\begin{equation}
    (1-\lambda)m_\lambda
    \le
   J_{\mathrm{avg}}^*
    \le
    (1-\lambda)M_\lambda,
    \qquad \forall \mathbf b_0\in\B .
    \label{eq:g_interval_app}
\end{equation}
Moreover, by definition of \(m_\lambda\) and \(M_\lambda\),
\begin{equation}
    (1-\lambda)m_\lambda
    \le
    (1-\lambda)V^*(\mathbf b)
    \le
    (1-\lambda)M_\lambda,
    \qquad \forall \mathbf b\in\B .
    \label{eq:scaled_value_interval_app}
\end{equation}
Therefore, both \(J_{\mathrm{avg}}^*\) and \((1-\lambda)V^*(\mathbf b)\) lie in
the same interval
\[
    \left[(1-\lambda)m_\lambda,\,(1-\lambda)M_\lambda\right].
\]
Consequently according to the Squeeze theorem,
\begin{equation}
    \left|
    (1-\lambda)V^*(\mathbf b)-J_{\mathrm{avg}}^*
    \right|
    \le
    (1-\lambda)(M_\lambda-m_\lambda),
\end{equation}
or equivalently,
\begin{equation}
    \left|
    (1-\lambda)V^*(\mathbf b)-J_{\mathrm{avg}}^*
    \right|
    \le
    (1-\lambda)\operatorname{span}(V^*(\mathbf{b})).
    \label{eq:value_level_bound_app}
\end{equation}
Hence
\begin{equation}
    J_{\mathrm{avg}}^*
    \ge
    J_\lambda^*(\mathbf b)
    -
    (1-\lambda)\operatorname{span}(V^*(\mathbf{b})),
\end{equation}
which completes the proof. It remains to explain how \(\operatorname{span}(\widehat V(\mathbf{b}))\) is computed from the PWLC representation. Let \(\mathcal G\) denote the finite set of supporting vectors returned by IPA, so that
\begin{equation}
    \widehat V(\mathbf b)
    =
    \min_{\bm\gamma\in\mathcal G}
    \bm\gamma^\top \mathbf b .
\end{equation}
Define
\begin{equation}
    \underline V
    :=
    \inf_{\mathbf b\in\mathcal B}\widehat V(\mathbf b),
    \qquad
    \overline V
    :=
    \sup_{\mathbf b\in\mathcal B}\widehat V(\mathbf b).
\end{equation}
Since \(\mathbf b\) lies in the probability simplex, for any fixed \(\bm\gamma\), the quantity \(\bm\gamma^\top\mathbf b\) is a convex combination of the entries of \(\bm\gamma\). Hence,
\begin{equation}
    \inf_{\mathbf b\in\mathcal B}
    \bm\gamma^\top \mathbf b
    =
    \min_i \gamma_i .
\end{equation}
Therefore,
\begin{equation}
    \underline V
    =
    \min_{\bm\gamma\in\mathcal G}
    \min_i \gamma_i .
\end{equation}
The upper endpoint \(\overline V\) is obtained by solving the following linear program:
\begin{equation}
\begin{aligned}
    \overline V
    =
    \max_{\mathbf b,\xi}\quad & \xi \\
    \text{s.t.}\quad
    & \xi \le \bm\gamma^\top \mathbf b,
    \qquad \forall \bm\gamma\in\mathcal G,\\
    & \mathbf 1^{\trans} \mathbf b = 1,\\
    & \mathbf b \ge 0 .
\end{aligned}
\end{equation}
Thus,
\begin{equation}
    \operatorname{span}(\widehat V)
    =
    \overline V-\underline V .
\end{equation}
This shows that the computable certificate is obtained from the finite vector set \(\mathcal G\), while its tightness depends on the quality and size of \(\mathcal G\) and discounted factor $\lambda$.

\bibliographystyle{IEEEtran}
\bibliography{Bib_References/IEEEabrv,Bib_References/conf_short,Bib_References/refer}
\end{document}